\documentclass[aps,prd,preprint,showpacs]{revtex4}
\usepackage{graphicx,amssymb,amsmath}

\begin{document}

\title{Reference priors for high energy physics}

\author{Luc Demortier}
\affiliation{Laboratory of Experimental High Energy Physics, Rockefeller University, New York, New York 10065}

\author{Supriya Jain}
\affiliation{Homer L. Dodge Department of Physics and Astronomy, University of Oklahoma, Norman, Oklahoma 73019}
\altaffiliation{Now at the State University of New York, Buffalo, New York 14260}

\author{Harrison B. Prosper}
\affiliation{Department of Physics, Florida State University, Tallahassee, Florida 32306}

\date{\today}

\begin{abstract}
Bayesian inferences in high energy physics often use uniform prior distributions 
for parameters about which little or no information is available before data are
collected.  The resulting posterior distributions are therefore sensitive to the 
choice of parametrization for the problem and may even be improper if this 
choice is not carefully considered.  Here we describe an extensively tested 
methodology, known as reference analysis, which allows one to construct 
parametrization-invariant priors that embody the notion of minimal 
informativeness in a mathematically well-defined sense.  We apply this 
methodology to general cross section measurements and show that it yields 
sensible results.  A recent measurement of the single top quark cross section 
illustrates the relevant techniques in a realistic situation.  
\end{abstract}

\pacs{06.20.Dk,14.65.Ha}

\maketitle

\section{Introduction}
The Bayesian approach~\cite{Robert2007} to inference plays an increasingly 
important role in particle physics research.  This is due, in part, to a better 
understanding of Bayesian reasoning within the field and the concomitant abating 
of the frequentist/Bayesian debate. Moreover, the small but growing number of 
successful applications provide concrete examples of how the Bayesian approach 
fares in practice.

In spite of these successes the specification of priors in a principled way 
remains a conceptual and practical hurdle.  In the so-called subjective Bayesian 
approach~\cite{OHagan1994}, one is invited to elicit the prior based on one's 
actual beliefs about the unknown parameters in the problem.  If one has 
well-understood information, for example based on subsidiary measurements or 
simulation studies, one can encode this partial information in an evidence-based 
prior~\cite{Cox2007}.  Such priors generally occasion little or no controversy.  
On the other hand, if one knows little about a given parameter, or if one 
prefers to act as if one knows little, then it is far from clear how one ought 
to encode this minimal information in a prior probability.  

Since there is, in fact, no unique way to model prior ignorance, a viewpoint
has evolved in which this lack of knowledge is represented by one's willingness 
to {\em adopt} a standard prior for certain parameters~\cite{Kass1996}, just as 
one has adopted a standard for quantities such as length and weight.  In this 
spirit, our field adopted as a convention a uniform (flat) prior for unknown 
cross sections and other parameters (see for example Ref.~\cite{Bertram2000}), 
mainly because this prescription is simple to implement and seems to embody 
Laplace's principle of insufficient reason. Unfortunately, uniform priors are 
both conceptually and practically flawed.  The conceptual difficulty is with 
their justification: lack of knowledge about a parameter $\sigma$ implies lack 
of knowledge about any one-to-one transform $\sigma^{\prime}$ of $\sigma$, and 
yet a prior distribution that is uniform in $\sigma$ will not be so in 
$\sigma^{\prime}$ if the transform is non-linear.  The practical problem is 
that careless use of uniform priors can lead to improper posteriors, that is, 
posteriors whose integrals are infinite and which can therefore not be used to 
assign meaningful probabilities to subsets of parameter space.  An example of
this pathology is found in a common method for reporting the exclusion of a new 
physics signal, where one estimates an upper limit from a posterior distribution
for the signal's production cross section.  When constructed from a Poisson 
probability mass function for the observations, a flat prior for the signal 
cross section, and a truncated Gaussian prior for the signal acceptance, this 
posterior is actually improper.  However, for small acceptance uncertainties 
the divergence of the upper limit is often concealed by the inevitable 
truncation of numerical computations~\cite{Demortier2002}.

The specification of priors that encode minimal information is of such 
importance in practice that a large body of literature exists describing 
attempts to construct priors that yield results with provably useful
characteristics. These priors are typically arrived at using formal rules.  
In this paper, therefore, we refer to them as {\em formal 
priors}~\cite{FormalPriors}  to distinguish them from evidence-based priors.  
Many such formal rules exist~\cite{Kass1996}.  In this paper we study, and 
then recommend, a rule which is arguably the most successful: that developed by 
Bernardo~\cite{Bernardo1979} and Berger and Bernardo~\cite{Berger1989,
Berger1992a,Berger1992b}.  Formal priors constructed according to the 
Bernardo-Berger rule are called {\em reference priors}, a somewhat unfortunate 
name given that the term reference prior is sometimes used as a synonym for 
what we have called a formal prior.
 
Reference priors have been shown to yield results with several desirable 
properties, all of which should appeal to particle physicists.  Therefore, in 
principle such priors could be a foundation for Bayesian inference in particle 
physics research.  However, reference priors and the associated methods 
collectively referred to as reference analysis~\cite{Bernardo2005,Demortier2005} 
have yet to enter the field in a significant way. The purpose of this paper is 
to initiate this process by applying the Bernardo-Berger method to a familiar, 
but important class of problems, namely that of calculating posterior densities 
for signal cross sections. 

In the next section we describe the general goals of reference prior 
construction and show how these are implemented via the concept of
missing information.  For simplicity we limit that discussion to one-parameter 
problems.  Section~\ref{InclPartInfo} then considers the treatment of nuisance 
parameters about which prior information is available.  Examples of such 
parameters include detector calibration constants, background contaminations, 
geometrical acceptances, and integrated luminosities.  We describe two methods 
for handling these parameters, depending on the type of information that is
available about them.  These methods are then applied to counting experiments 
with uncertain background contamination and effective luminosity.  In the
simplest cases we have obtained analytical expressions for the marginal 
posterior for the quantity of interest.  For the general case we have developed 
a numerical algorithm.  Some appealing properties of these posteriors are 
examined in Sec.~\ref{ValidationStudies}.  In Sec.~\ref{SingleTop} the 
reference prior methodology is applied to a recent measurement of the production 
cross section for single top quarks at the Tevatron.  Final comments are
presented in Sec.~\ref{FinalComments}.

\section{Reference Priors}
\label{sec:RefPriors}
In 1979, Bernardo~\cite{Bernardo1979} introduced a formal rule for constructing
what he called a reference prior. The goal was to construct a prior which, in a 
sense to be made precise, contained as little information as possible relative 
to the statistical model under consideration. By statistical model he meant a 
representation of the entire experimental design, including the probability 
distribution of the data, the sampling space, and the stopping rule.  Hence, by
construction reference priors depend on all these aspects of a statistical 
model, and so will inferences derived from data with the help of a reference
prior.  This may seem to violate the so-called {\em likelihood 
principle}~\cite{Birnbaum1962}, according to which all the information 
about unknown model parameters obtainable from an experiment is contained in the 
likelihood function, i.e. the probability distribution of the data, evaluated 
at the observations and viewed as a function of the parameters.  While this 
is formally true, it should be kept in mind that the likelihood principle
applies after data have been observed, whereas reference priors are constructed
at the experimental design stage.  Their purpose is to approximate a consensus
of opinions that is suitable for scientific communication.  This is generally
unproblematic in large-sample situations, where posterior inferences are 
dominated by the likelihood function.  In small sample cases however, results
obtained with reference priors should be considered preliminary, and a careful
study should be conducted of the degree to which inferences about the physics
model underlying the observations can be trusted.  This can be achieved by 
examining the sensitivity of the results to changes in the prior, and 
subsequently assessing the need for additional observations.

\begin{figure}
\includegraphics[width=15cm]{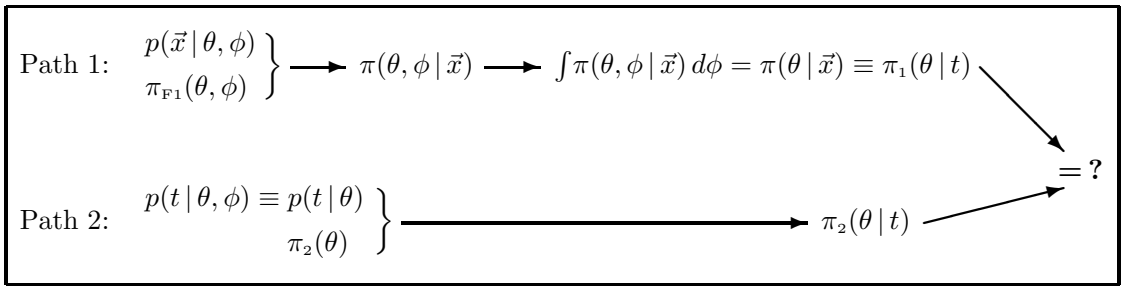}
\caption{Let $\vec{x}$ be a dataset modeled by the probability density 
$p(\vec{x}\,|\,\theta,\phi)$, where $\theta$ and $\phi$ are unknown parameters, 
and consider the following two paths to a posterior density for $\theta$.  In 
path~1, we use a formal prior $\pi_{\scriptscriptstyle\rm F1}(\theta,\phi)$ to
construct the joint posterior for $\theta$ and $\phi$, and then integrate out 
$\phi$.  Suppose that the result of this operation only depends on the data 
$\vec{x}$ through the statistic $t$; this gives us 
$\pi_{\scriptscriptstyle 1}(\theta\,|\,t)$.  For path~2, assume further that 
the sampling distribution of $t$ only depends on $\theta$.  We can then 
directly construct a posterior for $\theta$, say 
$\pi_{\scriptscriptstyle 2}(\theta\,|\,t)$.  A marginalization paradox occurs if 
$\pi_{\scriptscriptstyle 1}(\theta\,|\,t)\ne 
\pi_{\scriptscriptstyle 2}(\theta\,|\,t)$ regardless of the choice of 
prior $\pi_{\scriptscriptstyle 2}(\theta)$ in path~2.
\label{fig:mparadox}
}
\end{figure}

Reference priors have several desirable properties, including
\begin{enumerate}
\item {\em generality:} a well-defined algorithm exists to create a reference
      prior for almost any type of estimation problem, and the resulting 
      posterior is proper;
\item {\em invariance:} given a one-to-one map from a parameter $\theta$ to 
      a parameter $\phi$, applying the reference prior construction separately 
      to $\theta$ and $\phi$ yields posteriors that are related by the correct 
      transformation law, $\pi(\phi\,|\,x) = \pi(\theta\,|\,x) \,
      |\partial\theta/\partial\phi|$;
\item {\em sampling consistency:} the posterior densities from an ensemble of 
      experiments tend to cluster around the true values of the parameters; and
\item {\em coherence:} inferences derived from reference priors avoid 
      marginalization paradoxes.  
\end{enumerate}
Marginalization paradoxes~\cite{Dawid1973} arise in multiparameter problems 
when a posterior density can be calculated in different ways that ought to give 
the same answer but do not (see Fig.~\ref{fig:mparadox}).
This incoherence does not happen with subjective or evidence-based priors because 
these priors are always proper.  With formal priors however, it can only be 
avoided by allowing the joint prior for all the parameters in a given statistical 
model to depend on the quantity of interest.  This is in fact what the reference 
prior construction does.  For a simple illustration, consider $n$ measurements 
$x_{i}$ from the normal model with unknown mean $\mu$ and standard deviation 
$\sigma$.  The likelihood function is:
\begin{equation}
p(\vec{x}\,|\,\mu,\sigma)
\;=\; \prod_{i=1}^{n} \frac{e^{-\tfrac{1}{2}
      \bigl(\tfrac{x_{i}-\mu}{\sigma}\bigr)^{2}}}{\sqrt{2\pi}\,\sigma}
\;=\; \frac{e^{-\tfrac{n-1}{2}\bigl(\tfrac{s}{\sigma}\bigr)^{2}-
      \tfrac{n}{2}\bigl(\tfrac{\bar{x}-\mu}{\sigma}\bigr)^{2}}}
      {\bigl(\sqrt{2\pi}\,\sigma\bigr)^{n}},
\end{equation}
where $\bar{x}=\sum_{i}^{n}x_{i}/n$ and $s^{2}=\sum_{i}^{n}(x_{i}-\bar{x})^{2}
/(n-1)$.  When $\mu$ is the quantity of interest, the reference prior derived
from this likelihood is $1/\sigma$.  Restricting the remaining calculations to
the case $n=2$ for convenience, the joint reference posterior for $\mu$ and 
$\sigma$ is then:
\begin{equation}
\pi(\mu,\sigma\,|\,\vec{x})\;=\;\frac{\sqrt{2}\,s}{\pi\,\sigma^{3}}\;
   e^{-\tfrac{1}{2}\bigl(\tfrac{s}{\sigma}\bigr)^{2}-
   \bigl(\tfrac{\mu-\bar{x}}{\sigma}\bigr)^{2}}.
\label{eq:MuSigmaPost}
\end{equation}
Integrating out $\sigma$ yields the marginal $\mu$-posterior, which is a Cauchy
distribution with location parameter $\bar{x}$ and scale parameter $s/\sqrt{2}$.  
Suppose however that our interest lies in the standardized mean 
$\theta=\mu/\sigma$.  In a non-reference approach one would perform the 
transformation $(\mu,\sigma)\rightarrow (\theta,\sigma)$ in 
Eq.~\eqref{eq:MuSigmaPost} and integrate out $\sigma$ in order to obtain the 
marginal $\theta$-posterior.  The latter only depends on the data through the 
statistic $t\equiv \sqrt{2}\,\bar{x}/s$:
\begin{equation}
\pi(\theta\,|\,\vec{x})
   \;=\;\frac{e^{-\tfrac{\theta^{2}}{1+t^{2}}}}{\sqrt{\pi\,(1+t^{2})}}\,
        \left[1+\textrm{erf}\left(\frac{t\theta}{\sqrt{1+t^{2}}}\right)\right]
   \;=\;p(\theta\,|\,t),
\label{eq:phiPost}
\end{equation}
where $\textrm{erf}$ is the error function.  Furthermore, the sampling 
distribution of $t$ turns out to depend on $\theta$ only and is a noncentral 
Student's $t$ distribution for one degree of freedom and with noncentrality 
parameter $\theta$:
\begin{equation}
p(t\,|\,\theta)\;=\; \frac{e^{-\theta^{2}}}{\pi\,(1+t^{2})}\;+\;
\frac{\theta\,t\,e^{-\tfrac{\theta^{2}}{1+t^{2}}}}{\sqrt{\pi}\,(1+t^{2})^{3/2}}\;
\left[1 + \textrm{erf}\left(\frac{\theta\, t}{\sqrt{1+t^{2}}}\right)\right].
\label{eq:tSamplingDis}
\end{equation}
It is clear that there exists no prior (no function of $\theta$ only) that, 
multiplied by the likelihood~\eqref{eq:tSamplingDis}, leads to the 
posterior~\eqref{eq:phiPost}.  Hence the marginalization paradox: someone who 
is only given the data value of the statistic $t$ will be able to make 
inferences about the parameter $\theta$, but these inferences are guaranteed 
to disagree with those previously made by the Bayesian who had access to the 
full dataset.  Resolution of this paradox hinges on the realization that lack 
of information about $\mu$ is not the same as lack of information about 
$\theta$.  Therefore, the choice of which quantity is of interest must be 
done {\em  before} calculating the prior.  Since reference priors are derived 
from the likelihood function, the latter must first be expressed in terms of 
the relevant parameters, $\theta$ and $\sigma$:
\begin{equation}
p(\vec{x}\,|\,\theta,\sigma)
\;=\; \frac{e^{-\tfrac{n-1}{2}\bigl(\tfrac{s}{\sigma}\bigr)^{2}-
                \tfrac{n}{2}\bigl(\tfrac{\bar{x}}{\sigma}-\theta\bigr)^{2}}}
      {\bigl(\sqrt{2\pi}\,\sigma\bigr)^{n}}.
\end{equation}
Applying the reference algorithm to this likelihood while treating $\theta$ as 
the quantity of interest yields the prior $\pi(\theta,\sigma) = 
1/(\sigma\,\sqrt{1+\theta^{2}/2})$, which is very different from the prior 
$1/\sigma$ obtained by treating $\mu$ as the quantity of interest.  The 
resulting reference posterior suffers no marginalization problems.  Further 
details about this example can be found in Refs.~\cite{Bernardo1979,Bernardo2005}.

Our discussion of marginalization also helps to clarify the behavior of 
reference posteriors under transformations in the multiparameter setting.  
Reference posteriors are invariant under one-to-one transformations of the 
parameter of interest, but not under transformations that redefine the parameter 
of interest by mixing in one or more nuisance parameters.  However, redefining 
the nuisance parameters is permitted.  Suppose for example that $\varphi$ is the 
parameter of interest, $\nu$ the nuisance parameter(s), and consider an 
invertible transformation of the form $(\varphi,\nu)\rightarrow 
(\varphi,\lambda)$, where $\lambda$ is a function of both $\varphi$ and $\nu$.  
Then the reference posterior for $\varphi$ is unchanged by the transformation.

Reference priors on unbounded parameter spaces are usually improper, which
invalidates the application of Bayes' theorem.  To circumvent this problem one 
introduces a nested sequence of compact subsets $\Theta_{1}\subset\Theta_{2}
\subset\ldots$ of the parameter space $\Theta$, such that $\Theta_{\ell}
\rightarrow\Theta$ as $\ell\rightarrow\infty$.  Given an improper prior 
$\pi(\theta)$, its restriction to $\Theta_{\ell}$ will be proper, so that Bayes' 
theorem can be applied to construct the corresponding restricted posterior 
$\pi_{\ell}(\theta|x)$.  The unrestricted posterior for the entire parameter
space is then defined by the limit of the $\pi_{\ell}(\theta|x)$ as 
$\ell\rightarrow\infty$.  The practical justification for this procedure is 
that one often knows the shape, but not the size, of the physical region of 
parameter space where the prior has nonzero weight.  As this size is typically 
very large, the limiting posterior can be viewed as an approximation to the 
posterior on the physical region.  

Interestingly, the limiting posterior can also be obtained by direct, formal 
application of Bayes' theorem to the improper prior $\pi(\theta)$, provided the 
marginal distribution of the data,
\begin{equation}
m(x)\;\equiv\; \int p(x|\theta)\, \pi(\theta) \,d\theta,
\end{equation}
is finite.  It can then be shown that the restricted posteriors 
$\pi_{\ell}(\theta|x)$ converge {\em logarithmically} to their limit 
$\pi(\theta|x)$:
\begin{equation}
\lim_{\ell\rightarrow\infty} D[\pi_{\ell}(\theta|x),\,\pi(\theta|x)]\;=\;0,
\end{equation}
where
\begin{equation}
D[p(\theta),\,q(\theta)]\;\equiv\;
\int\!q(\theta)\,\log\frac{q(\theta)}{p(\theta)}\,d\theta
\end{equation}
is the Kullback-Leibler divergence between $p(\theta)$ and $q(\theta)$.  This 
divergence is a parametrization-independent, non-negative measure of the 
separation between two densities; it is zero if and only if the densities are 
identical. Unfortunately, pointwise logarithmic convergence is not enough to 
avoid inferential inconsistency in some special cases~\cite{Berger2009}, so 
that a stronger form of convergence is needed, {\em expected} logarithmic 
convergence:
\begin{equation}
\lim_{\ell\rightarrow\infty} 
\mathbb{E}_{m}\Bigl\{D[\pi_{\ell}(\theta|x),\,\pi(\theta|x)]\Bigr\}\;=\;0,
\end{equation}
where the expectation is taken with respect to the marginal density $m(x)$.

The above discussion motivates the following terminology~\cite{Berger2009}.  
Given a statistical model on a parameter space $\Theta$, a {\em standard prior} 
is a strictly positive and continuous function on $\Theta$ that yields a proper 
posterior.  A {\em permissible prior} is a standard prior for which the 
posterior is the expected logarithmic limit of a sequence of posteriors defined
by restriction to compact sets.

\subsection{The Concept of Missing Information}
Reference priors make use of the notion of expected intrinsic information.  
For one observation from a model $p(x|\theta)$, the expected intrinsic 
information about the value of $\theta$ when the prior is $\pi(\theta)$ is 
given by the functional
\begin{equation}
I\{\pi\} \;=\; \mathbb{E}_{m}\Bigl\{ D[\pi(\theta|x),\;\pi(\theta)] \Bigr\}.
\label{eq:xii}
\end{equation}
The more informative the observation, the greater the expected separation 
between the posterior and the prior.  The larger this separation, the greater 
the expected intrinsic information $I\{\pi\}$.  Thus, $I\{\pi\}$ measures the 
amount of information about the value of $\theta$ that might be expected from 
one observation when the prior is $\pi(\theta)$.  

Suppose next that we make $k$ independent observations $x_{(k)}=\{x_{1}, x_{2},
\ldots, x_{k}\}$ from the model $p(x|\theta)$.  The definition of expected
intrinsic information can be generalized to include all $k$ observations:
\begin{equation}
I_{k}\{\pi\} \;=\; 
             \mathbb{E}_{m}\Bigl\{ D[\pi(\theta|x_{(k)}),\;\pi(\theta)] \Bigr\},
\label{eq:xii2}
\end{equation}
where the expectation is a $k$-dimensional integral over $x_{(k)}$ weighted by:
\begin{equation}
m(x_{(k)})\;\equiv\; \int p(x_{(k)}|\theta)\, \pi(\theta) \,d\theta
          \;=\; \int \left[\prod_{i=1}^{k}p(x_{i}|\theta)\right]\;\pi(\theta)\,d\theta.
\end{equation}
As the sample size $k$ grows larger, one expects the amount of information about 
$\theta$ to increase, and in the limit $k\rightarrow\infty$, the true value of 
$\theta$ would become exactly known.  In this sense, the limit $I_{\infty}\{\pi\}
\equiv\lim_{k\rightarrow\infty}I_{k}\{\pi\}$ represents the {\em missing} information 
about $\theta$ when $\pi(\theta)$ is the prior.  This concept of missing 
information is central to the construction of reference priors.

\subsection{Reference Priors for One-Parameter Models}
\label{sec:refprior}
The goal of reference analysis is to contruct a prior that maximizes the missing 
information.  This maximization cannot be done directly however, because 
$I_{\infty}$ typically diverges.  To avoid this problem, one first constructs 
the prior $\pi_{k}(\theta)$ that maximizes $I_{k}$, and then takes the limit of 
$\pi_{k}(\theta)$ as $k\rightarrow\infty$.  Additional care is required when 
the parameter space $\Theta$ is unbounded, since in that case the prior that 
maximizes $I_{k}$ is often improper, and $I_{k}$ is undefined for improper 
priors.  The solution is to define reference priors via their restrictions on 
arbitrary compact subsets $\Theta_{\ell}$ of $\Theta$.  Thus one is led to the 
formal definition of a reference prior for $\theta$ as any permissible prior 
$\pi_{R}(\theta)$ that satisfies the so-called maximizing missing information 
(MMI) property, namely that
\begin{equation}
\lim_{k\rightarrow\infty}\Bigl[ I_{k}\{\pi_{R,\ell}\}-I_{k}\{\pi_{\ell}\}\Bigr]
\;\ge\;0
\label{eq:refdef}
\end{equation}
for any compact set $\Theta_{\ell}$ and candidate prior $\pi(\theta)$, where 
$\pi_{R,\ell}$ and $\pi_{\ell}$ are the renormalized restrictions of $\pi_{R}$ 
and $\pi$ to $\Theta_{\ell}$.  A candidate prior is a standard prior that 
incorporates any prior knowledge about $\theta$.  

A key result is the following constructive definition of the reference prior 
$\pi_{R}(\theta)$~\cite{Berger2009},
\begin{eqnarray}
\pi_{R}(\theta) \; & =  & \; \lim_{k\rightarrow\infty}
                       \frac{\pi_{k}(\theta)}{\pi_{k}(\theta_{0})}, \nonumber \\
\mbox{with} \; \pi_k(\theta)   \; &  =  & \; \exp\left\{ \int p(x_{(k)}\,|\,\theta)\,
                \ln\left[\frac{p(x_{(k)}\,|\,\theta)\, h(\theta)}
                {\int p(x_{(k)}\,|\,\theta)\, h(\theta)\, d\theta}\right]\,
                dx_{(k)}\right\},
\label{eq:refformula}
\end{eqnarray}
where $\theta_{0}$ is an arbitrary fixed point in $\Theta$, $h(\theta)$ is any 
continuous, strictly positive function, such as $h(\theta)=1$, and 
$p(x_{(k)}\,|\,\theta) = \prod_{i=1}^k p(x_i\,|\,\theta)$ is the probability 
model for a sample of $k$ independent observations.  We emphasize that this 
constructive definition only guarantees that the MMI property~\eqref{eq:refdef} 
is satisfied.  The permissibility part of the reference prior definition must be 
separately verified.  However, the proponents of reference priors view the MMI 
property as considerably more important than permissibility~\cite{Berger2009}, 
and also believe that it would be highly unusual for a prior satisfying the MMI 
property to fail permissibility~\cite{Berger1992b} (counter-examples are known,
but they are rather exotic).

A further useful result that we shall exploit is that, when certain regularity 
conditions are met --- essentially those that guarantee asymptotic normality of 
the posterior --- the reference prior for models with one continuous parameter 
reduces to the well-known Jeffreys prior~\cite{Kass1996},
\begin{equation}
\pi_{R}(\theta)\;=\;\sqrt{\mathbb{E}\left[-\frac{d^{2}}{d\theta^{2}}
                    \ln p(x\,|\,\theta)\right]},
\label{eq:Jpr}
\end{equation}
where the expectation is taken with respect to the sampling model 
$p(x\,|\,\theta)$.  In general the analytical derivation of reference priors 
can be extremely challenging.  However, Eq.~\eqref{eq:refformula} is amenable to 
numerical integration~\cite{Berger2009}.  

We emphasize that the definition and results described in this section apply 
only to the case where the model of interest depends on a single parameter.  
A generalization to the multi-parameter case has been formulated and shown to 
have the properties listed at the beginning of 
Sec.~\ref{sec:RefPriors}~\cite{Bernardo2005}.  
We shall not describe it here however, except for when evidence-based priors are 
specified for the additional parameters.  This is a very common situation in 
high energy physics, and will be discussed next.

\section{Nuisance Parameters}
\label{InclPartInfo}
The reference prior algorithm described in Sec.~\ref{sec:refprior} pertains to 
models containing no nuisance parameters. In practice, however, every non-trivial
problem must contend with such parameters and the reference prior algorithm 
must be generalized accordingly. In this paper we restrict our attention to 
nuisance parameters for which partial information is available, which is often 
the case in practice. 

Depending on the type of partial information that is available, there are two 
plausible ways one might choose to incorporate nuisance parameters $\phi$ into 
the calculation of the reference priors for a parameter of interest 
$\theta$~\cite{Sun1998}:  
\begin{description}
\item[Method 1:] Assume that we are given a marginal prior $\pi(\phi)$ for the
                 nuisance parameters;  compute the conditional reference 
                 prior $\pi_{R}(\theta\,|\,\phi)$ for the interest parameter given 
                 a fixed value of $\phi$;  the full prior is then  
                 $\pi(\theta, \phi) = \pi_{R}(\theta\,|\,\phi)\, \pi(\phi)$;
\item[Method 2:] Assume that we are given a conditional prior 
                 $\pi(\phi\,|\,\theta)$ for the nuisance parameter given the 
                 interest parameter;  marginalize the probability model 
                 $p(x|\theta, \phi)$ with respect to $\phi$ in order to obtain 
                 $p(x|\theta)=\int p(x|\theta,\phi)\,\pi(\phi|\theta)\,d\phi$, 
                 and compute the reference prior $\pi_{R}(\theta)$ for the 
                 marginalized model;  the full prior is then 
                 $\pi(\theta,\phi) = \pi(\phi\,|\,\theta)\,\pi_{R}(\theta)$.
\end{description}
In many high energy physics measurements there are often sound reasons for assuming
that the nuisance parameter is independent of the parameter of interest.  
Information about a detector energy scale, for example, is typically determined
separately from the measurement of interest, say of a particle mass, and
is therefore considered to be independent \emph{a priori}  from one's information about
the particle's mass.  When an experimenter is willing to make this assumption,
he or she can declare that $\pi(\phi\,|\,\theta) = \pi(\phi)$ and use Method 2.
When this assumption does not seem fully justified, and it is too difficult to
elicit the $\theta$ dependence of $\pi(\phi\,|\,\theta)$, then it will seem
preferable to use Method 1, which only requires knowledge of the marginal 
prior $\pi(\phi)$.  When one is unsure of which method to use, one should use 
both, and treat the results as part of a test of robustness.  An important
practical advantage of Method 1 is that the conditional reference prior is 
computed once and for all, for a given model, and can be used with any 
evidence-based prior for the nuisance parameters.  In contrast, for Method 2 
the reference prior must be computed anew every time the priors for the 
nuisance parameters change.  On the other hand, since Method 2 reduces the problem
to one involving a single parameter, the reference prior algorithm reduces to 
Jeffreys' rule~\eqref{eq:Jpr}, which is typically easier to implement.

In the next section we introduce the basic model studied in this paper, and
follow with the application of Methods~1 and~2 to that model. 

\subsection{The Single-Count Model} 
A very common model for high energy physics measurements is the following.  
A number of events $N$ is observed by some apparatus, and it is assumed that 
$N$ is Poisson distributed with mean count $\epsilon\,\sigma+\mu$, where 
$\sigma$ is the rate of a physics signal process, typically the cross section, 
which we detect with an effective integrated luminosity $\epsilon$ --- that is, 
the integrated luminosity scaled by the signal efficiency, and $\mu$ is a 
background contamination.  Thus, $\sigma$ is the parameter of interest, whereas 
$\epsilon$ and $\mu$ are nuisance parameters for which we usually have partial 
information. For physical reasons none of these three parameters can be 
negative.  We write the likelihood for this model as
\begin{equation}
p(n|\sigma,\epsilon,\mu)\;=\; \frac{(\epsilon\sigma+\mu)^{n}}{n!}\;
                                         e^{-\epsilon\sigma-\mu}
\quad\textrm{with}\;\; 0\le \sigma <\infty \;\;
\textrm{and}\;\; 0<\epsilon, \mu <\infty.
\label{eq:scm}
\end{equation}
Information about $\epsilon$ and $\mu$ usually comes from a variety of sources, 
such as auxiliary measurements, Monte Carlo simulations, theoretical 
calculations, and evidence-based beliefs (for example, some sources of 
background contributing to $\mu$ may be deemed small enough to ignore, and some 
physics effects on $\epsilon$, such as gluon radiation, may be believed to be 
well enough reproduced by the simulation to be reliable ``within a factor of 
2'').  It is therefore natural to represent that information by an evidence-based 
prior.  Here we will assume that $\epsilon$ and $\mu$ are independent of 
$\sigma$ and that their prior factorizes as a product of two gamma densities:
\begin{equation}
\pi(\epsilon,\mu\,|\,\sigma)\;=\;\pi(\epsilon,\mu)\;=\;
   \frac{a(a\epsilon)^{x-1/2}\,e^{-a\epsilon}}{\Gamma(x+1/2)}\;
   \frac{b(b\mu)^{y-1/2}\,e^{-b\mu}}{\Gamma(y+1/2)},
\label{eq:emuprior}
\end{equation}
where $a$, $b$, $x$, and $y$ are known constants, related to the means
$\bar{\epsilon}$, $\bar{\mu}$ and coefficients of variation $\delta\epsilon$,
$\delta\mu$ by:
\begin{equation}
\bar{\epsilon} = \frac{x+\frac{1}{2}}{a},\quad
\delta\epsilon = \frac{1}{\sqrt{x+\frac{1}{2}}},\quad
\bar{\mu}      = \frac{y+\frac{1}{2}}{b},\quad
\delta\mu      = \frac{1}{\sqrt{y+\frac{1}{2}}}.
\label{eq:PriorParameters}
\end{equation}
The built-in assumption that $\epsilon$ and $\mu$ are uncorrelated is clearly 
an approximation, since they share a dependence on the integrated luminosity, 
which is itself uncertain.

There are two ways of interpreting this prior.  The first one is appropriate 
when information about $\epsilon$ and $\mu$ comes from one or more 
non-experimental sources, such as Monte Carlo studies and theoretical 
calculations, and takes the form of a central value plus an uncertainty.  Since 
the $\epsilon$ and $\mu$ components of the prior are each modeled by a 
two-parameter density, one can fix the shape of this density in each case by 
matching its mean with the central value of the corresponding measurement and 
its standard deviation with the uncertainty.  It will then be necessary to 
check the robustness of the final analysis results to reasonable changes in this 
procedure.  For example, one may want to replace the gamma distribution by a 
log-normal or truncated Gaussian one, and the mean by the mode or median.  

The second interpretation of prior~\eqref{eq:emuprior} follows from the analysis
of two independent, auxiliary Poisson measurements, in which the observed number
of events is $x$ for the effective luminosity and $y$ for the background.  The
expected numbers of events in these auxiliary measurements are $a\epsilon$ and
$b\mu$, respectively.  For a Poisson likelihood with mean $a\epsilon$ the 
reference prior coincides with Jeffreys' prior and is proportional to
$1/\sqrt{\epsilon}$.  Given a measurement $x$, the posterior will then be a
gamma distribution with shape parameter $x+1/2$ and scale parameter $1/a$.  A
similar result holds for the background measurement.  In this manner the 
prior~\eqref{eq:emuprior} is obtained as a joint reference posterior from two 
auxiliary measurements.

The problem we are interested in is finding a prior for $\sigma$, about which 
either little is known or one wishes to act as if this is so.

\subsubsection{Application of Method 1 to the Single-Count Model}
\label{Method1SingleCount}
This section serves two purposes: to illustrate the analytical algorithm for 
computing reference priors and to apply Method~1 to model~\eqref{eq:scm}.

In Method 1~\cite{Sun1998}, we find first the conditional reference prior 
$\pi_{R}(\sigma\,|\,\epsilon,\mu)$ and then multiply by the evidence-based prior
$\pi(\epsilon,\mu)$ to construct the full prior $\pi(\sigma,\epsilon,\mu)$.  As 
will be illustrated in Sec.~\ref{ValidationStudies}, the single-count model 
is regular enough to warrant using Jeffreys' rule in the first step of the 
calculation of $\pi_{R}(\sigma\,|\,\epsilon,\mu)$.  We therefore apply 
Eq.~\eqref{eq:Jpr} to the $\sigma$ dependence of the likelihood~\eqref{eq:scm}, 
while holding $\epsilon$ and $\mu$ constant; this yields:
\begin{equation}
\pi_{J}(\sigma\,|\,\epsilon,\mu) 
    \;\propto\; \sqrt{\mathbb{E}\left[-\frac{\partial^{2}}{\partial\sigma^{2}}
    \ln p(n|\sigma,\epsilon,\mu)\right]}
    \;\propto\; \frac{\epsilon}{\sqrt{\epsilon\,\sigma+\mu}}.
\label{eq:piJ}
\end{equation}
This prior is clearly improper with respect to $\sigma$ and is therefore only 
defined up to a proportionality constant.  However, this constant could very 
well depend on $\epsilon$ and $\mu$, since we kept these parameters fixed in 
the calculation.  It is important to obtain this dependence correctly, as
examples have shown that otherwise inconsistent Bayes estimators may result.
Reference~\cite{Sun1998} proposes a compact subset normalization procedure.  
One starts by choosing a nested sequence $\Theta_{1}\subset\Theta_{2}\subset
\cdots$ of compact subsets of the parameter space $\Theta = \{(\sigma,\epsilon,
\mu)\}$, such that $\cup_{\ell}\Theta_{\ell} = \Theta$ and the integral 
$K_{\ell}(\epsilon,\mu)$ of $\pi_{J}(\sigma\,|\,\epsilon,\mu)$ over 
$\Omega_{\ell}\equiv\{\sigma: (\sigma,\epsilon,\mu)\in\Theta_{\ell}\}$ is 
finite.  The conditional reference prior for $\sigma$ on $\Omega_{\ell}$ is then
\begin{equation}
\pi_{R,\ell}(\sigma\,|\,\epsilon,\mu)
  \;=\;  \frac{\pi_{J}(\sigma\,|\,\epsilon,\mu)}{K_{\ell}(\epsilon,\mu)}.
\label{eq:piRell}
\end{equation}
To obtain the conditional reference prior on the whole parameter space, one
chooses a fixed point $(\sigma_{0},\epsilon_{0},\mu_{0})$ within that space 
and takes the limit of the ratio
\begin{equation}
\pi_{R}(\sigma\,|\,\epsilon,\mu)\;\propto\; \lim_{\ell\rightarrow\infty}
\frac{\pi_{R,\ell}(\sigma\,|\,\epsilon,\mu)}
     {\pi_{R,\ell}(\sigma_{0}\,|\,\epsilon_{0},\mu_{0})}.
\label{eq:piR}
\end{equation}
By taking the limit in this ratio form, one avoids problems arising from 
$K_{\ell}(\epsilon,\mu)$ becoming infinite as $\ell\rightarrow\infty$.

The theory of reference priors currently does not provide guidelines for 
choosing the compact sets $\Theta_{\ell}$, other than to require that the 
resulting posterior be proper.  In most cases this choice makes no difference
and one is free to base the choice of compact sets on considerations of 
simplicity and convenience.  However, we have found that some care is required 
with the single-count model.  Indeed, suppose we make the plausible choice
\begin{equation}
\Theta_{\ell} \;=\; \Bigl\{(\sigma,\epsilon,\mu):\; \sigma\in[0,u_{\ell}],\;
\epsilon\in[0,v_{\ell}],\; \mu\in[0,w_{\ell}]\Bigr\},
\label{eq:cset1}
\end{equation}
where $\{u_{\ell}\}$, $\{v_{\ell}\}$, and $\{w_{\ell}\}$ are increasing 
sequences of positive constants.  If we use these sets in applying 
Eqs.~\eqref{eq:piRell} and~\eqref{eq:piR} to the prior~\eqref{eq:piJ}, 
we obtain:
\begin{equation}
\pi_{R}(\sigma\,|\,\epsilon,\mu)
 \;\propto\;\sqrt{\frac{\epsilon}{\epsilon\,\sigma+\mu}}.
\label{eq:piR1}
\end{equation}
Although this prior is still improper with respect to $\sigma$, its dependence 
on $\epsilon$ is different from that of the conditional Jeffreys' prior, 
Eq.~\eqref{eq:piJ}.  This demonstrates the potential importance of the 
compact subset normalization.  The prior in Eq.~\eqref{eq:piR1} has a serious 
problem however.  Suppose that the $\epsilon$ marginal of our evidence-based 
prior for $\epsilon$ and $\mu$ is $\exp(-\epsilon)/\sqrt{\pi\epsilon}$.  It is 
then easy to verify that the resulting posterior is improper, since its 
$\epsilon$ marginal has the non-integrable form $\exp(-\epsilon)/\epsilon$.  
The cause of this problem is the choice of compact sets~\eqref{eq:cset1}.  

Fortunately it is not difficult to find a sequence of compact sets that will
provide a proper posterior.  Indeed, the $\sigma$ dependence of the 
prior~\eqref{eq:piJ} suggests that the compact sets should be based on
the parametrization $(\epsilon\sigma,\epsilon,\mu)$ rather than
$(\sigma,\epsilon,\mu)$~\cite{BergerPC2008}.  We therefore set:
\begin{equation}
\Theta_{\ell} \;=\; \Bigl\{(\sigma,\epsilon,\mu):\; 
\sigma\in[0,u_{\ell}/\epsilon],\;\epsilon\in[1/v_{\ell},v_{\ell}],\; 
\mu\in[0,w_{\ell}]\Bigr\},
\label{eq:NCS2}
\end{equation}
where $u_{\ell}$, $v_{\ell}$, and $w_{\ell}$ are as before.  Again using
Eqs.~\eqref{eq:piJ}, \eqref{eq:piRell}, and~\eqref{eq:piR}, we now find:
\begin{equation}
\pi_{R1}(\sigma\,|\,\epsilon,\mu)\;\propto\;
\frac{\epsilon}{\sqrt{\epsilon\sigma+\mu}},
\label{eq:piR1b}
\end{equation}
which is identical to Jeffreys' prior for this problem and yields 
well-behaved posteriors.  For future use, the subscript $R1$ on the left-hand 
side indicates that this reference prior was obtained with Method~1.

We now have all the ingredients needed to calculate the marginal reference
posterior $\pi_{R1}(\sigma\,|\,n)$ for the cross section $\sigma$: the 
likelihood~\eqref{eq:scm}, the marginal nuisance prior~\eqref{eq:emuprior}, 
and the conditional reference prior~\eqref{eq:piR1b}.  For calculating 
posterior summaries in terms of intervals and upper limits it is convenient
to express the result as a tail probability:
\begin{equation}
\int_{\sigma}^{\infty}\!\!\pi_{R1}(\tau\,|\,n)\;d\tau\;=\;
\int_{\!\frac{\sigma}{a+\sigma}}^{1}
\frac{u^{n+y}\;(1-u)^{x-\frac{1}{2}}}{B\bigl(n+y+1,x+\frac{1}{2}\bigr)}\;
\frac{B_{\frac{b}{b+1}(1+\frac{u-1}{u}\frac{\sigma}{a})}
      \bigl(y+\frac{1}{2},n+\frac{1}{2}\bigr)}
     {B_{\frac{b}{b+1}}\bigl(y+\frac{1}{2},n+\frac{1}{2}\bigr)}
\;du
\end{equation}
where
\begin{equation}
B_{z}(u,v)\;\equiv\;\int_{0}^{z}\! t^{u-1}\;(1-t)^{v-1}\;dt
\end{equation}
is the incomplete beta function, and $B(u,v)\equiv B_{1}(u,v)\;=\;
\Gamma(u)\Gamma(v)/\Gamma(u+v)$.

\subsubsection{Application of Method 2 to the Single-Count Model} 
In contrast with Method~1, Method~2 requires from the start that we specify the 
evidence-based prior for the effective integrated luminosity $\epsilon$ and the 
background contamination~$\mu$.  Furthermore, this specification must be done
conditionally on the signal rate $\sigma$.  As mentioned earlier, we will use
expression~\eqref{eq:emuprior} for this prior.

The next step in the application of Method~2 is to marginalize the probability
model~\eqref{eq:scm} with respect to $\epsilon$ and $\mu$:
\begin{align}
p(n\,|\,\sigma) 
            &\;=\;\iint p(n\,|\,\sigma,\epsilon,\mu)\;
                  \pi(\epsilon,\mu\,|\,\sigma)\;d\epsilon\,d\mu,\nonumber\\[2mm]
            &\;=\;\iint \frac{(\epsilon\sigma+\mu)^{n}}{n!}\;
                  e^{-\epsilon\sigma-\mu}\;
                  \frac{a(a\epsilon)^{x-\frac{1}{2}}}{\Gamma(x+\frac{1}{2})}\;
                  e^{-a\epsilon}\;
                  \frac{b(b\mu)^{y-\frac{1}{2}}}{\Gamma(y+\frac{1}{2})}\;
                  e^{-b\mu}\;d\epsilon\,d\mu,\nonumber\\[2mm]
            &\;=\;\left[\frac{a}{a+\sigma}\right]^{x+\frac{1}{2}}\;
                  \left[\frac{b}{b+1}\right]^{y+\frac{1}{2}}\; S_{n}^{0}(\sigma),
\label{eq:MarginalModel}
\end{align}
where 
\begin{equation}
S_{n}^{m}(\sigma)\;\equiv\;
\sum_{k=0}^{n} k^{m}\;
\binom{k+x-\frac{1}{2}}{k}\;\binom{n-k+y-\frac{1}{2}}{n-k}\;
\left[\frac{1}{b+1}\right]^{n-k}\;
\left[\frac{\sigma}{a+\sigma}\right]^{k},
\end{equation}
and the binomial coefficients are expressed in terms of gamma functions to 
accomodate noninteger values of their arguments.  Finally, the reference prior 
algorithm must be applied to the marginalized model $p(n\,|\,\sigma)$.  As in 
the case of Method 1, the conditions for applying Jeffreys' rule are satisfied 
here; we therefore obtain:
\begin{equation}
\pi_{R2}(\sigma)\;\propto\;\sqrt{\sum_{n=0}^{\infty}
\frac{\bigl[(x+\frac{1}{2})\,S_{n}^{0}(\sigma)\,-\,
            \frac{a}{\sigma}\,S_{n}^{1}(\sigma)\bigr]^{2}}
     {(a+\sigma)^{x+5/2}\;S_{n}^{0}(\sigma)}}.
\label{eq:piR2}
\end{equation}
We will use the notation $\pi_{R2}(\sigma)$ to refer to the marginal reference 
prior for $\sigma$ obtained with Method~2.  Note that the compact subset argument
invoked in the construction of the Method~1 reference prior is not needed here
because all the parameters other than $\sigma$ have already been eliminated by
marginalization.  

For Method~2 the marginal reference posterior for $\sigma$ is proportional to
the product of the marginal data probability distribution~\eqref{eq:MarginalModel}
and the marginal reference prior~\eqref{eq:piR2}:
\begin{equation}
\pi_{R2}(\sigma\,|\,n)\;\propto\;p(n\,|\,\sigma)\;\pi_{R2}(\sigma).
\end{equation}
The normalization of $\pi_{R2}(\sigma\,|\,n)$ must be obtained numerically. 

\subsection{The Multiple-Count Model}
An important generalization of the single-count model is obtained by considering
$M$ replications of the latter; the likelihood is:
\begin{equation}
p(\vec{n}\,|\,\sigma,\vec{\epsilon},\vec{\mu})\;=\;
\prod_{i=1}^{M}\frac{(\epsilon_{i}\sigma+\mu_{i})^{n_{i}}}{n_{i}!}\,
e^{-\epsilon_{i}\sigma-\mu_{i}}.
\end{equation}
To obtain the Method~1 reference prior for this model, we first calculate
Jeffreys' prior for $\sigma$, while keeping $\vec{\epsilon}$ and $\vec{\mu}$
fixed:
\begin{equation}
\pi_{J}(\sigma\,|\,\vec{\epsilon},\vec{\mu})\;\propto\;
\sqrt{\sum_{i=1}^{M}\frac{\epsilon_{i}^{2}}{\epsilon_{i}\,\sigma+\mu_{i}}}.
\end{equation}
This prior is improper, requiring us to apply the compact subset normalization
described in Sec.~\ref{Method1SingleCount}.  Using a straightforward
generalization of the nested compact sets of Eq.~\eqref{eq:NCS2}, we find that
the correct reference prior is identical to Jeffreys' prior.

In order to apply Method~2, we need to specify a proper conditional prior for 
the $\mu_{i}$ and $\epsilon_{i}$ given $\sigma$.  Neglecting correlations, we
set:
\begin{equation}
\pi(\vec{\epsilon},\vec{\mu}|\sigma)\;=\;
\prod_{i=1}^{M} \frac{a_{i}(a_{i}\epsilon_{i})^{x_{i}-1/2}\,
                e^{-a_{i}\epsilon_{i}}}{\Gamma(x_{i}+1/2)}\;
                \frac{b_{i}(b_{i}\mu_{i})^{y_{i}-1/2}\,
                e^{-b_{i}\mu_{i}}}{\Gamma(y_{i}+1/2)}.
\end{equation}
The marginalized data probability distribution $p(\vec{n}|\sigma)$ is then a
product of expressions of the form~\eqref{eq:MarginalModel}, one for each count 
$i$.  

Here we no longer attempt to obtain analytical expressions for the Method~1 
and~2 reference posteriors.  Instead, we use the numerical algorithms described
below.

\subsection{Numerical Algorithms}
\label{NumericalAlgorithm}
In this section we describe numerical algorithms that can be used to compute 
Method~1 or~2 reference posteriors for the single- and multiple-count Poisson 
likelihoods discussed in the previous sections.  

For Method~1 the algorithm starts by generating $(\sigma,\vec{\epsilon},
\vec{\mu})$ triplets from the ``flat-prior posterior'', i.e. the posterior 
obtained by setting $\pi(\sigma\,|\,\vec{\epsilon},\vec{\mu})=1$ (line~3 in the 
pseudo-code below);  the correct reference prior $\pi(\sigma\,|\,\vec{\epsilon},
\vec{\mu})$ is then computed at lines~4--7 and is used at line~9 to weight the 
generated $\sigma$ values so as to produce the reference posterior:
{\small
\begin{tabbing}
~~~~~~~~\=~~~~~~~~\=~~~~~~~~\=~~~~~~~~\=\kill
\>{\scriptsize 1}\> Set $\vec{n}_{o}$ to the array of observed event numbers.\\
\>{\scriptsize 2}\> For $i=1,\ldots,I$:\\
\>{\scriptsize 3}\>\>Generate $(\sigma_{i},\vec{\epsilon}_{i},\vec{\mu}_{i}) \sim 
                     p(\vec{n}_{o}\,|\,\sigma,\vec{\epsilon},\vec{\mu})\,
                     \pi(\vec{\epsilon},\vec{\mu})$. \\
\>{\scriptsize 4}\>\> For $j=1,\ldots,J$:\\
\>{\scriptsize 5}\>\>\>Generate $\vec{n}_{j}\sim p(\vec{n}\,|\,\sigma_{i},
                       \vec{\epsilon}_{i},\vec{\mu}_{i})$.\\
\>{\scriptsize 6}\>\>\>Calculate $d^{2}[-\ln p(\vec{n}_{j}\,|\,\sigma_{i},
                       \vec{\epsilon}_{i},\vec{\mu}_{i})]/d\sigma_{i}^{2}$
                       by numerical differentiation.\\
\>{\scriptsize 7}\>\>Average the $J$ values of 
                     $d^{2}[-\ln p(\vec{n}\,|\,\sigma_{i},\vec{\epsilon}_{i},
                     \vec{\mu}_{i})]/d\sigma_{i}^{2}$ obtained\\
\>\>\>               at line 6, and take the square root.  This yields a 
                     numerical\\
\>\>\>               approximation to the conditional Jeffreys' prior 
                     $\pi_{J}(\sigma_{i}\,|\,\vec{\epsilon}_{i},\vec{\mu}_{i})$.\\
\>{\scriptsize 8}\> Histogram the $\sigma_{i}$ values generated at line 3, 
                    weighting them by\\
\>\>                $\pi_{J}(\sigma_{i}\,|\,\vec{\epsilon}_{i},\vec{\mu}_{i})/
                    p(\vec{n}_{o}\,|\,\sigma_{i},\vec{\epsilon}_{i},\vec{\mu}_{i})$.
                    This yields $\pi_{R1}(\sigma)$, the $\sigma$-marginal 
                    prior.\\
\>{\scriptsize 9}\> Histogram the $\sigma_{i}$ values generated at line 3, 
                    weighting them by \\
\>\>                $\pi_{J}(\sigma_{i}\,|\,\vec{\epsilon}_{i},\vec{\mu}_{i})$.
                    This yields $\pi_{R1}(\sigma\,|\,\vec{n}_{o})$, the 
                    $\sigma$-marginal posterior.\\
\end{tabbing}
}
Although not required for the calculation of the reference posterior, an
approximation to the reference prior is provided at line~8.  By construction 
this approximation is only reliable for $\sigma$ values in the bulk of the 
flat-prior posterior.  The generation step at line~3 is done via a Markov chain 
Monte Carlo procedure~\cite{BAT}.  The particular choice of sampling 
distribution for the generated $(\sigma,\vec{\epsilon},\vec{\mu})$ triplets is 
motivated by the desire to obtain weights with reasonably small variance at 
steps 8 and 9.  However, the flat-prior posterior $p(\vec{n}_{0}\,|\,\sigma,
\vec{\epsilon},\vec{\mu})\,\pi(\vec{\epsilon},\vec{\mu})$ is not always proper 
with respect to $(\sigma,\vec{\epsilon},\vec{\mu})$.  When $M=1$ (single-count 
model), it is improper if $x\le 1/2$.  Propriety can then be restored by 
multiplying the flat-prior posterior by $\epsilon$ and correspondingly
adjusting the weights at steps 8 and 9.  Another feature of the above algorithm 
is that it does not implement the compact subset normalization.  In the cases 
that we examined, this procedure made no difference, but this may not be true 
for more general problems than those our code seeks to solve.  Unfortunately 
the current lack of guidelines in the choice of compact sets limits our ability 
to address this issue in the code.

The algorithm for Method~2 has a simpler structure, since all it does is 
apply Jeffreys' rule to a marginalized likelihood $p(\vec{n}_{o}\,|\,\sigma)$
provided by the user.  The calculation does not require random sampling of the
parameters and is done at fixed $\sigma$ values.  For a given $\sigma$, the 
reference prior $\pi_{R2}(\sigma)$ is obtained by Monte Carlo averaging, over
an ensemble of vectors $\vec{n}$ generated from $p(\vec{n}\,|\,\sigma)$, of 
an accurate numerical approximation of the second derivative of the negative
log-likelihood~\cite{2derivative}.  As already pointed out, Method 2 does not
require a compact subset normalization procedure.  The reference posterior is
thus proportional to the product of $p(\vec{n}_{o}\,|\,\sigma)$ and 
$\pi_{R2}(\sigma)$, and the normalization with respect to $\sigma$ must be 
determined numerically.

\section{Validation Studies}
\label{ValidationStudies}
\begin{figure}
\includegraphics[width=7cm]{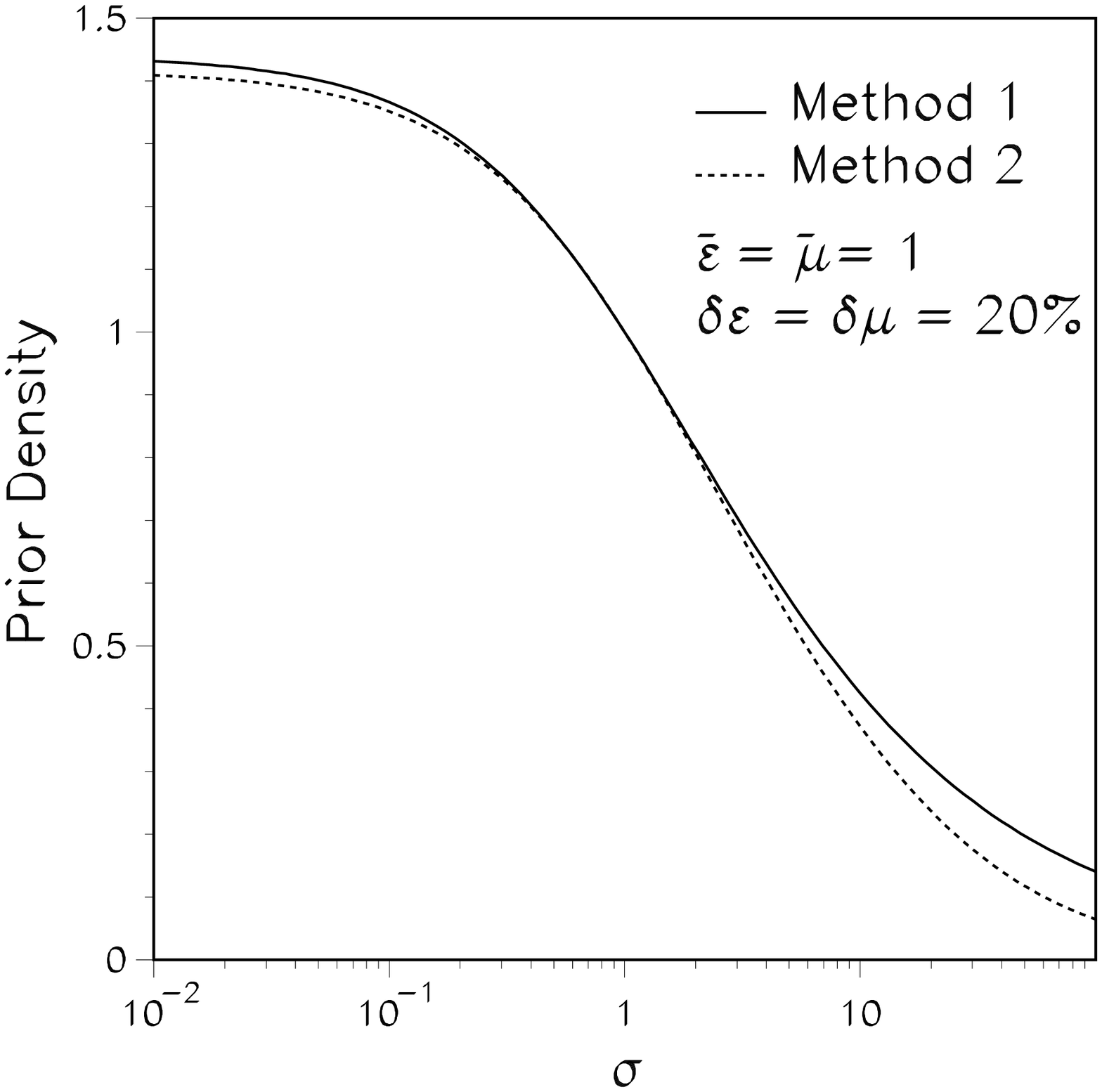}
\hspace*{1cm}
\includegraphics[width=7cm]{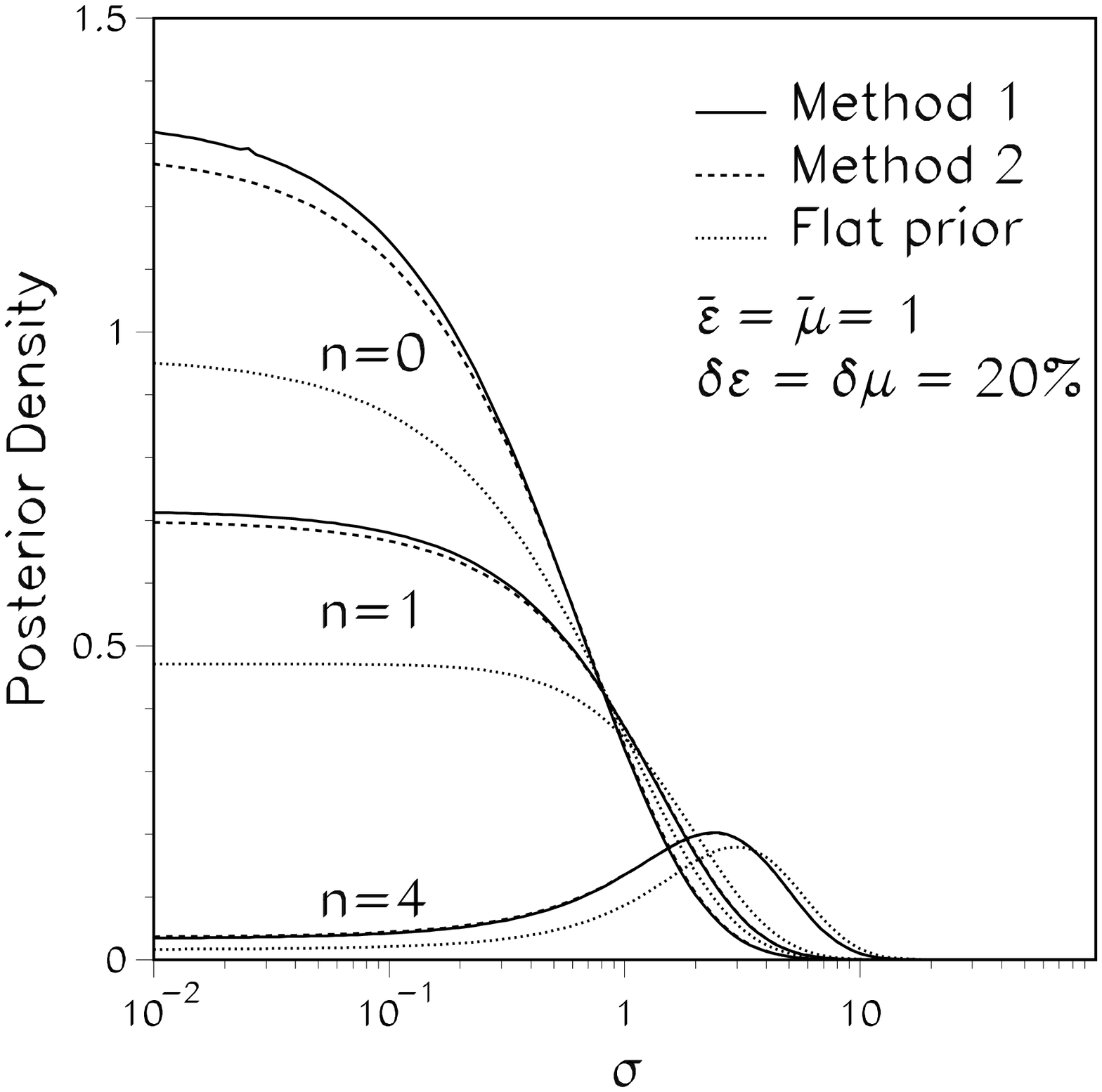}
\caption{Left: marginal Method-1 and 2 priors, normalized to 1 at $\sigma=1$.  
Right: marginal Method-1 and 2 posteriors for 0, 1, and 4 observed events,
together with the posteriors obtained from a flat prior.  The $\epsilon$ and 
$\mu$ priors have a mean of 1 and a 20\% coefficient of variation 
(corresponding to $x=y=24.5$ and $a=b=25$ in 
Eq.~\protect\eqref{eq:PriorParameters}).  Here and in subsequent plots, the 
units of $\sigma$ are arbitrary but consistent with those of $\epsilon$; 
e.g., if the latter is expressed in pb$^{-1}$, then $\sigma$ is given in pb so 
that, like $\mu$, the product of $\epsilon$ and $\sigma$ is dimensionless. 
\label{fig:priorMethod12}}
\end{figure}

We have performed a number of studies to validate inferences from the 
single-count model, using both the numerical algorithms described in 
Sec.~\ref{NumericalAlgorithm} and analytical expressions we obtained for the 
marginal Method-1 and~2 posteriors for $\sigma$.  To recapitulate, we have two 
reference priors for this model:
\begin{align}
\pi_{R1}(\sigma,\epsilon,\mu) &\;=\; \pi_{R1}(\sigma\,|\,\epsilon,\mu)\;
                                     \pi(\epsilon,\mu)\\[2mm]
\pi_{R2}(\sigma,\epsilon,\mu) &\;=\; \pi_{R2}(\sigma)\;
                                     \pi(\epsilon,\mu\,|\,\sigma),
\end{align}
and we have assumed that $\pi(\epsilon,\mu\,|\,\sigma)=\pi(\epsilon,\mu)$ at 
Eq.~\eqref{eq:emuprior}.  As explained in Sec.~\ref{InclPartInfo}, this 
extra assumption affects only the definition of $\pi_{R2}$, which therefore 
incorporates more information than $\pi_{R1}$.  In the present section we study 
and compare the properties of these two reference priors.  To begin, we show 
some example prior and posterior $\sigma$ marginals in 
Fig.~\ref{fig:priorMethod12}.
As expected, posteriors corresponding to a small observed number of events 
favor small cross sections, and posteriors derived from flat priors put less 
weight on small cross sections than reference posteriors.  

Our derivations of the two reference prior methods made use of Jeffreys' rule
\eqref{eq:Jpr}.  As pointed out in Sec.~\ref{sec:refprior}, this approach
assumes that some regularity conditions are satisfied, such that the resulting
posterior is asymptotically normal.  We now wish to verify this assumption with 
a graphical example.  If one adopts the objective Bayesian view that the 
parameters $\sigma$, $\epsilon$, and $\mu$ have true values, then the asymptotic 
limit can be defined as the result of a large number $N_{R}$ of replications of 
the  measurement, in the limit where that number goes to infinity.  For the case 
where each measurement replication $i$ consists of a number of events $n_{i}$ 
drawn from a probability mass function $f(n\,|\,\sigma,\epsilon,\mu)$, the 
reference posterior has the form:
\begin{equation}
\pi_{R}(\sigma,\epsilon,\mu\,|\,n_{1},n_{2},\ldots,n_{N_{R}}) \;\propto\;
\pi_{R}(\sigma,\epsilon,\mu)\,
\prod_{i=1}^{N_{R}}f(n_{i}\,|\,\sigma,\epsilon,\mu),
\label{eq:ReplicatedExp}
\end{equation}
and the reference prior $\pi_{R}(\sigma,\epsilon,\mu)$ is calculated from the 
combined likelihood for the $N_{R}$ measurements; it can also be calculated 
from a single one of these likelihood functions, since it follows from their 
constructive definition~\eqref{eq:refformula} that reference priors are 
independent of sample size.  For Method 1 the prior is given by the product of
Eqs.~\eqref{eq:emuprior} and~\eqref{eq:piR1b}, and the likelihood component 
$f(n\,|\,\sigma,\epsilon,\mu)$ by Eq.~\eqref{eq:scm}.  Replicating the 
measurement $N_{R}$ times is then equivalent to making a single measurement 
with a Poisson likelihood whose mean is $N_{R}$ times the original mean, and 
whose observation is the sum of the $N_{R}$ original observations $n_{i}$.  
This property of Poisson measurements simplifies the calculations considerably.  
For Method~2 the prior is given by Eq.~\eqref{eq:piR2} and the likelihood by 
Eq.~\eqref{eq:MarginalModel}.  In this case no simplification obtains when 
considering multiple replications, and numerical calculations must use 
explicitly the full product of likelihood functions.  
\begin{figure}
\includegraphics[width=7cm]{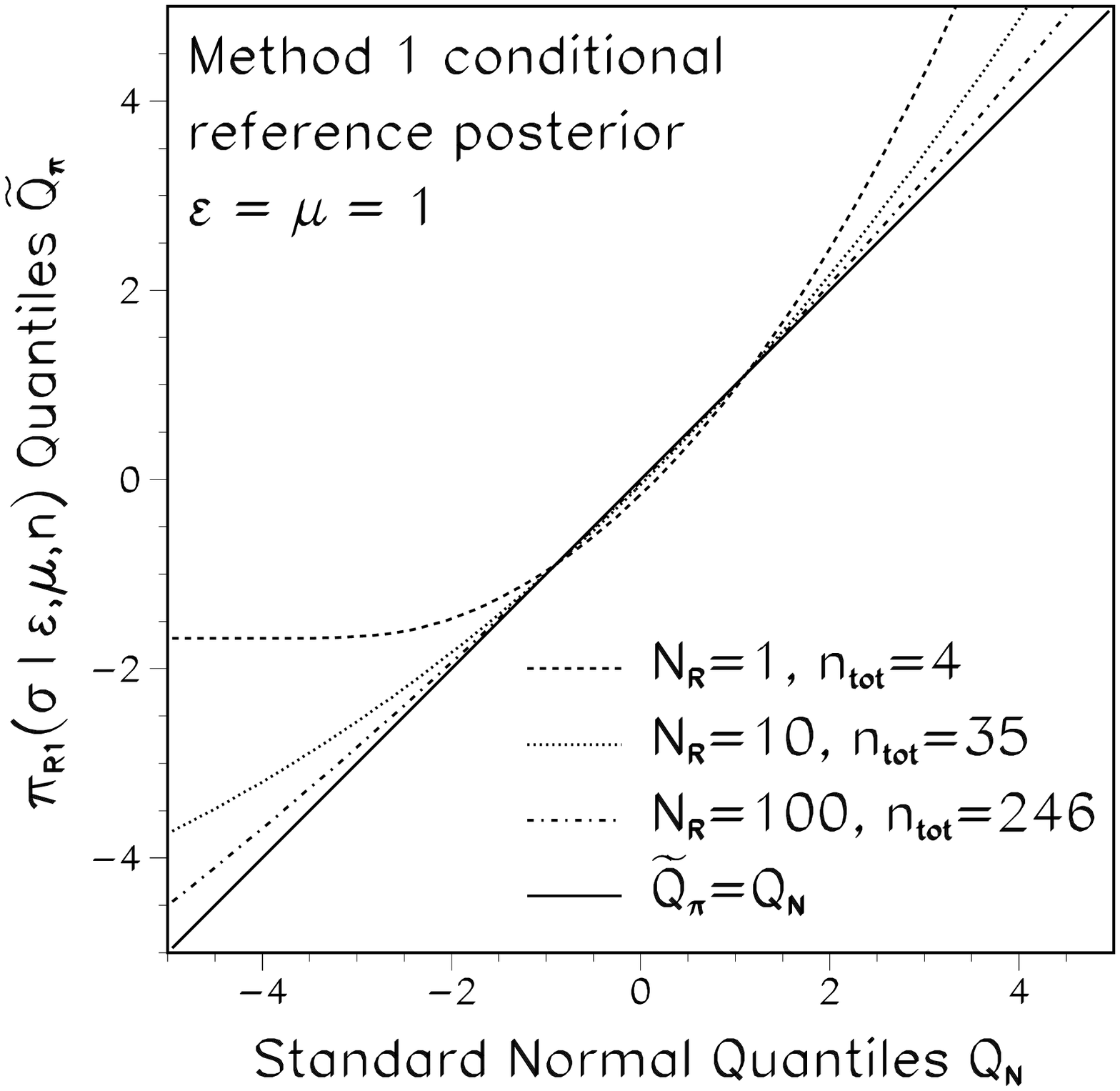}
\hspace*{1cm}
\includegraphics[width=7cm]{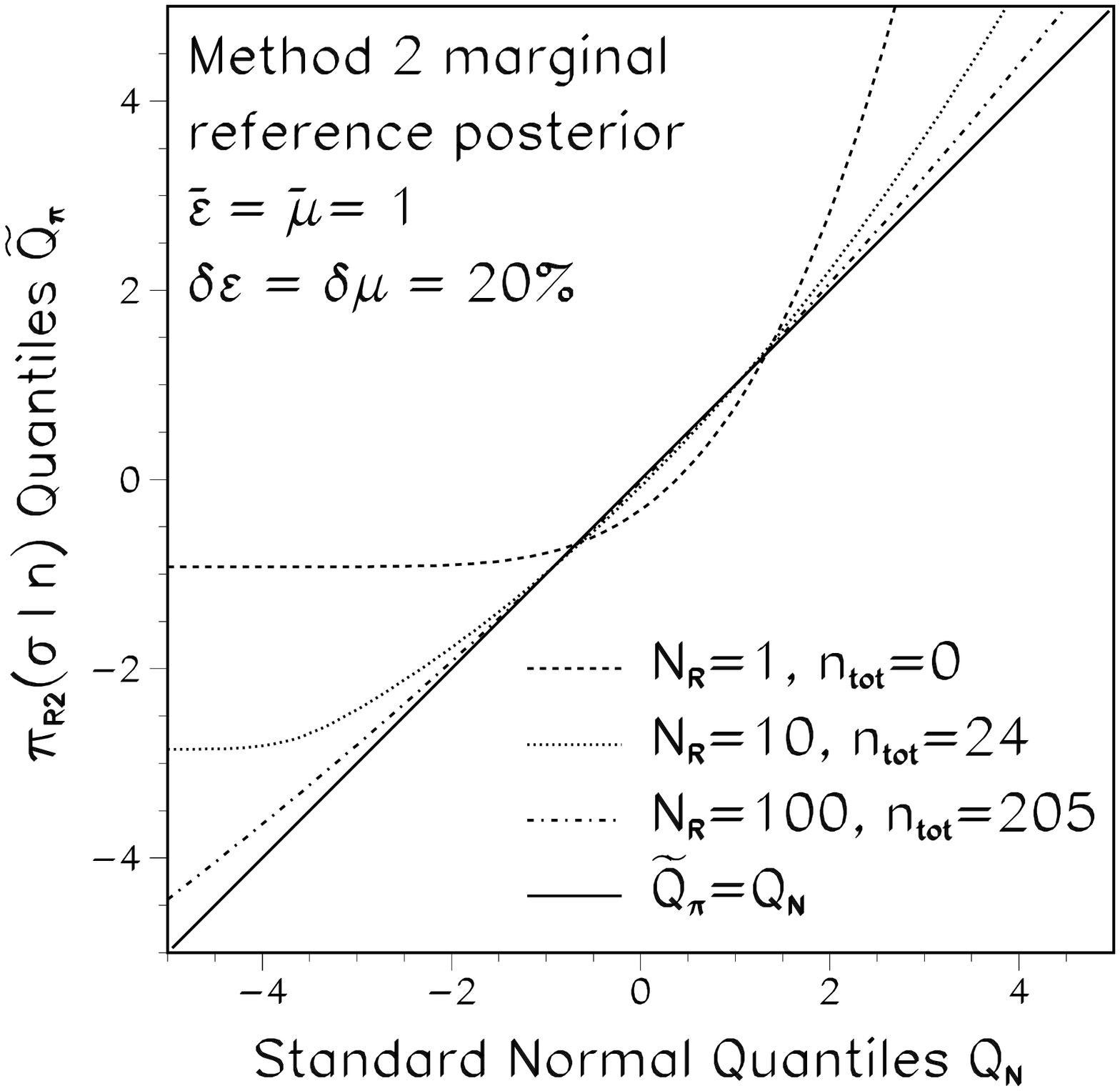}
\caption{Q-Q plots of the Method-1 conditional reference posterior (left) and
the Method-2 marginal reference posterior (right) versus the standard normal
distribution.  The reference posterior quantiles have been recentered (see 
text).  $N_{R}$ is the number of measurement replications and $n_{tot}$ is the
observed number of events summed over all replications.
\label{fig:m12qq}}
\end{figure}
Figure~\ref{fig:m12qq} illustrates the calculations with the help of so-called 
Q-Q plots, where recentered quantiles from the reference posterior for $\sigma$ 
are plotted against standard normal quantiles.  The posterior quantiles 
$Q_{\pi}$ are recentered according to 
$\tilde{Q}_{\pi}=(Q_{\pi}-\langle\sigma\rangle)/\Delta\sigma$,
where the posterior mean $\langle\sigma\rangle$ and standard deviation 
$\Delta\sigma$ are numerically estimated.  For Method~1 we set the true values
of $\sigma$, $\epsilon$, and $\mu$ to $1$.  We then randomly generate a 
sequence of 100 independent measurements from the probability mass 
function~\eqref{eq:scm} and use the subsequences with $N_{R}=1$, $10$, and 
$100$ to produce the curves in the left panel.  For Method~2 we set the true 
value of $\sigma$ to 1 and give the priors for $\epsilon$ and $\mu$ each a 
mean of 1 and a coefficient of variation of 20\%.  Measurements are then 
generated from the probability mass function~\eqref{eq:MarginalModel} in order 
to compute the curves in the right panel.  Both panels clearly show that the 
respective reference posteriors approach a Gaussian shape as the number of 
measurement replications increases.

Given the almost negligible difference between Method-1 and 2 posteriors 
exhibited in Fig.~\ref{fig:priorMethod12}, and the fact that our analytical 
results for Method~1 are computationally more tractable than those for Method~2, 
our considerations in the remainder of this section will focus exclusively on 
Method~1.

Among the reference prior properties listed in Sec.~\ref{sec:RefPriors}, the
ones of generality, invariance, and coherence are true by construction.  The
property of sampling consistency needs more elaboration however, since Bayesian 
inferences do not generally coincide with exact frequentist ones, and a proper 
evaluation requires first of all the specification of an ensemble of 
experiments.  A well-known property of Bayesian posterior intervals constructed
from a proper prior is that their coverage is exact when averaged over the 
prior~\cite{Pratt1965}.  This is an immediate consequence of the law of total 
probability.  Indeed, given a parameter $\theta$ with proper prior 
$\pi(\theta)$, and a measurement $X$, the prior-averaged frequentist coverage 
of a $1-\alpha$ Bayesian credibility interval $R(X)$ can be written as:
\begin{equation}
\mathbb{E}_{\pi}[\mathbb{P}(\theta\in R(X)\,|\,\theta)]
\,=\,\mathbb{P}(\theta\in R(X))
\,=\,\mathbb{E}_{m}[\mathbb{P}(\theta\in R(X)\,|\,X)]
\,=\,\mathbb{E}_{m}(1-\alpha)\,=\,1-\alpha,
\end{equation}
where the first expectation is over the prior $\pi(\theta)$ and the second one
over the marginal sampling distribution $m(x)=\int p(x|\theta)\,\pi(\theta)\:d\theta$.
When $\pi(\theta)$ is a reference prior, and especially when it is 
improper, there is no natural metric over which the coverage can be averaged.
The only sensible approach in that case is to study the coverage pointwise, 
i.e. as a function of the true value of $\theta$.  Since the single- and 
multiple-count models discussed in this paper combine an improper prior for the
parameter of interest $\sigma$ with proper priors for the nuisance parameters
$\vec{\epsilon}$ and $\vec{\mu}$, we will study interval coverage for a fixed
value of $\sigma$, but averaged over $\pi(\vec{\epsilon},\vec{\mu})$.  Our
interest is in how this coverage evolves toward the asymptotic limit.  As 
before, we take this limit in the sense of an ever-increasing number $N_{R}$ 
of experiment replications.  For a given value of $N_{R}$, the posterior is 
formed as in Eq.~\eqref{eq:ReplicatedExp} and its coverage is computed.  
\begin{figure}
\includegraphics[width=7cm]{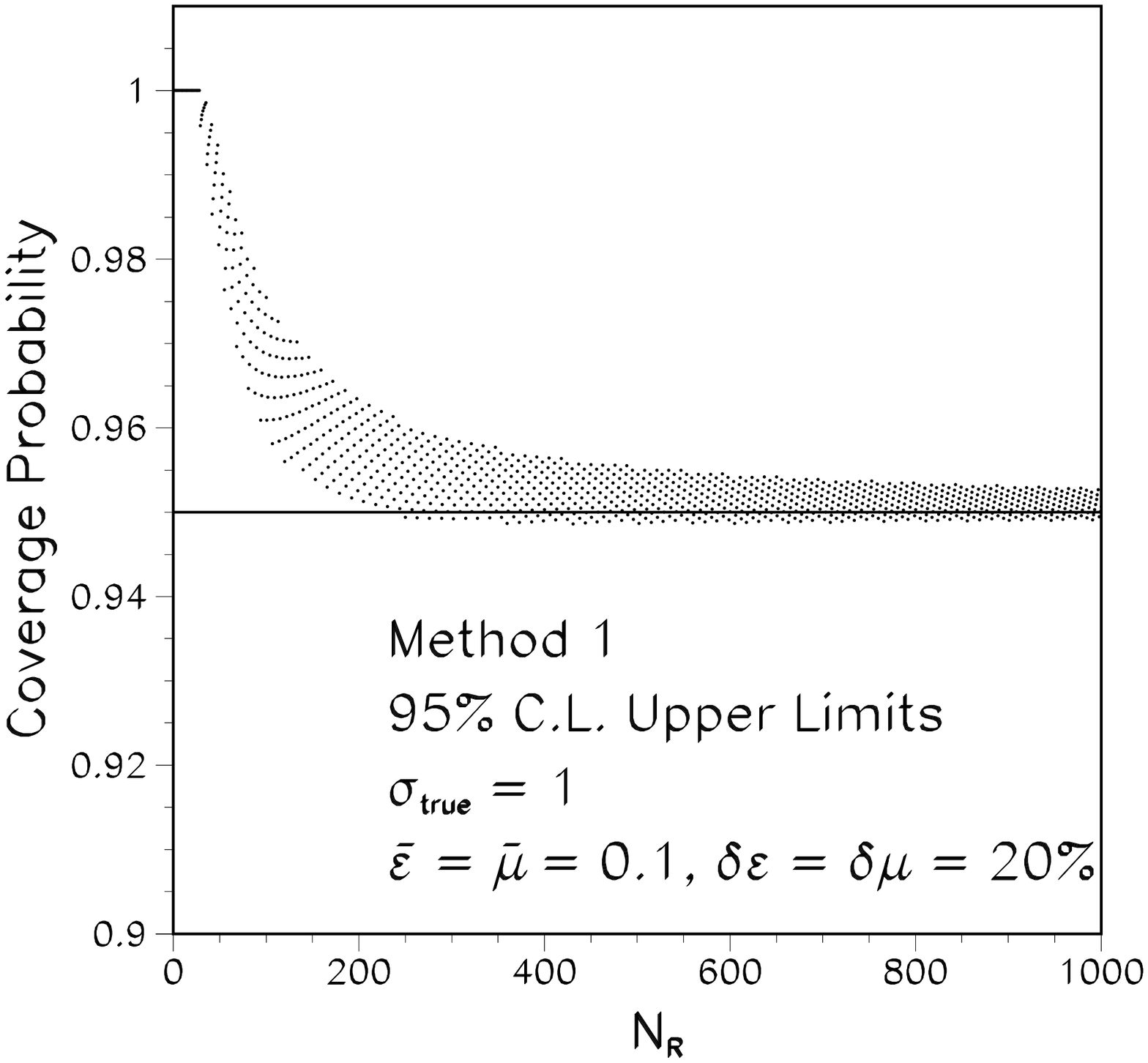}
\hspace*{1cm}
\includegraphics[width=7cm]{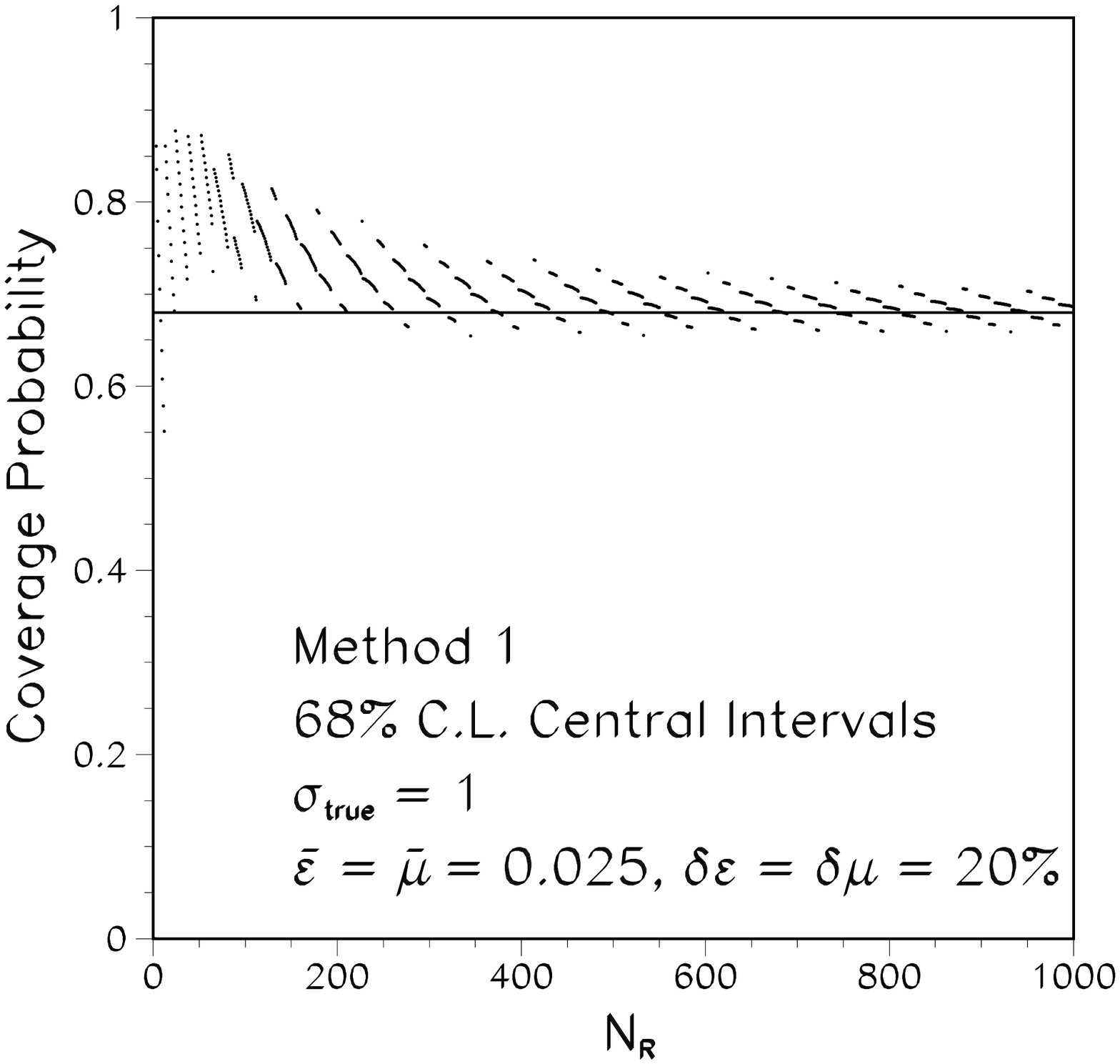}
\caption{Coverage probability of Method-1 posterior credibility upper limits 
(left) and central intervals (right), as a function of the number of experiment
replications $N_{R}$.  The solid lines indicate the nominal credibility.
\label{fig:m1coverage_xl}}
\end{figure}
Figure~\ref{fig:m1coverage_xl} shows the coverage of 95\% credibility upper 
limits and 68\% credibility central intervals as a function of $N_{R}$.  As 
the latter increases, the coverage converges to the credibility, confirming 
the sampling consistency of the method.

\begin{figure}
\includegraphics[width=7cm]{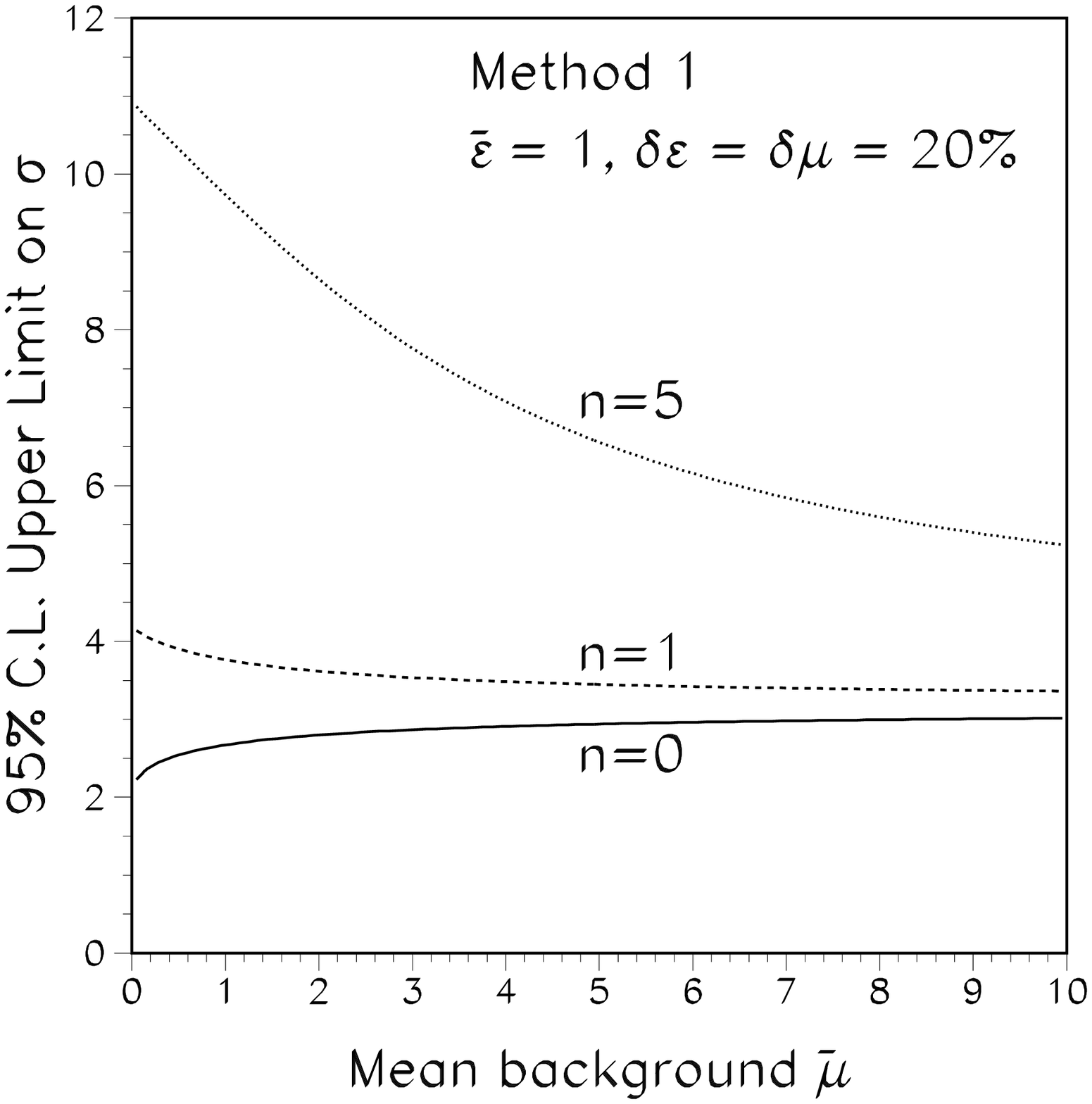}
\hspace*{1cm}
\includegraphics[width=7cm]{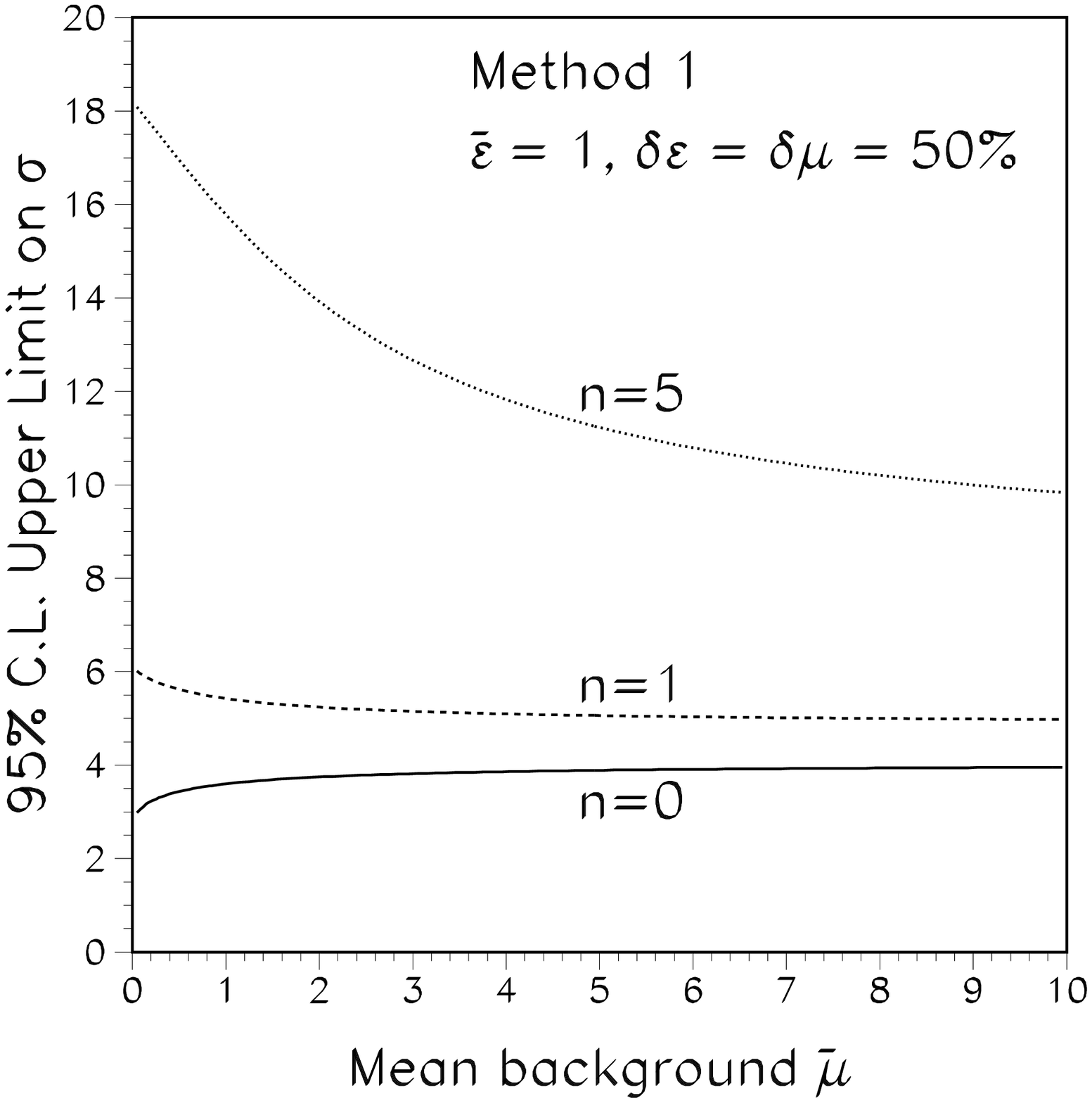}
\caption{Variation of the Method 1 reference posterior upper limit with mean 
background for several values of the observed number of events $n$.  The 
relative uncertainty on the background and on the effective luminosity is 20\% 
for the left plot and 50\% for the right one.
\label{fig:m1ul_meanbg}}
\end{figure}

Finally, we examine the behavior of reference posterior upper limits on $\sigma$
as a function of the expected background $\bar{\mu}$ (defined in 
Eq.~\eqref{eq:PriorParameters}) when the observed number of 
events $n$ is small.  For comparison, when $n=0$ and there are no uncertainties 
on signal efficiency and background, frequentist upper limits decrease 
linearly with background.  From a Bayesian point of view this result is 
surprising.  Indeed, when zero events are observed the likelihood function 
factorizes exactly into background and signal components, indicating that the 
experiment {\em actually} performed can be analyzed as the combination of two 
independent experiments, one to measure background and the other signal.  
If, in addition, signal and background are {\em a priori} independent, then 
posterior inferences about signal will be independent of background.  In 
particular, upper limits on $\sigma$ will be constant as a function of 
$\bar{\mu}$, not linearly decreasing.  The reference priors entangle signal and 
background however, so that upper limits will not be exactly constant.  The 
$n=0$ case is illustrated in Fig.~\ref{fig:m1ul_meanbg} for two values of the 
relative uncertainties on background and signal efficiency.
For $n>0$ the likelihood function still factorizes approximately since 
$(\mu+\epsilon\sigma)^{n}\approx \mu^{n} (1+n\epsilon\sigma/\mu)\approx\mu^{n}$ 
for $\mu\gg \epsilon\sigma, n$.  Thus upper limits will flatten out at large 
$\bar{\mu}$, as seen in Fig.~\ref{fig:m1ul_meanbg}.  A comparison of the left 
and right panels in that figure also shows that upper limits increase with the 
uncertainty on background and signal efficiency, as expected.

\section{Measurement of the Single Top Quark Cross Section}
\label{SingleTop}
In this section, we demonstrate the computational feasibility of the methods 
described above by applying them to the recent measurement of the single top 
cross section by the D0 and CDF collaborations~\cite{D0singletop,CDFsingletop}. 
Both collaborations use the same form of likelihood function --- a product of 
Poisson distributions over multiple bins of a multivariate discriminant, the
same form of evidence-based priors, namely truncated Gaussians, and flat priors 
for the cross section~\cite{MultipleBinFlatPrior}.  As a realistic example, 
we construct the reference prior for the cross section using one of the data 
channels considered by D0.

D0 partitioned their data into 24 channels, defined by lepton flavor (electron 
or muon), jet multiplicity (two, three, or four), number of $b$-tagged jets 
(one or two), and two data collection periods.  The discriminant distribution 
is shown in Fig.~3 of Ref.~\cite{D0singletop}. Here we consider the electron, 
two-jet, single-tag channel from one of the data taking periods.  The 
discriminant distribution contains about 500 counts spread over 50 bins,
with a maximum bin count of about 40.

\begin{figure}
\includegraphics[width=7cm]{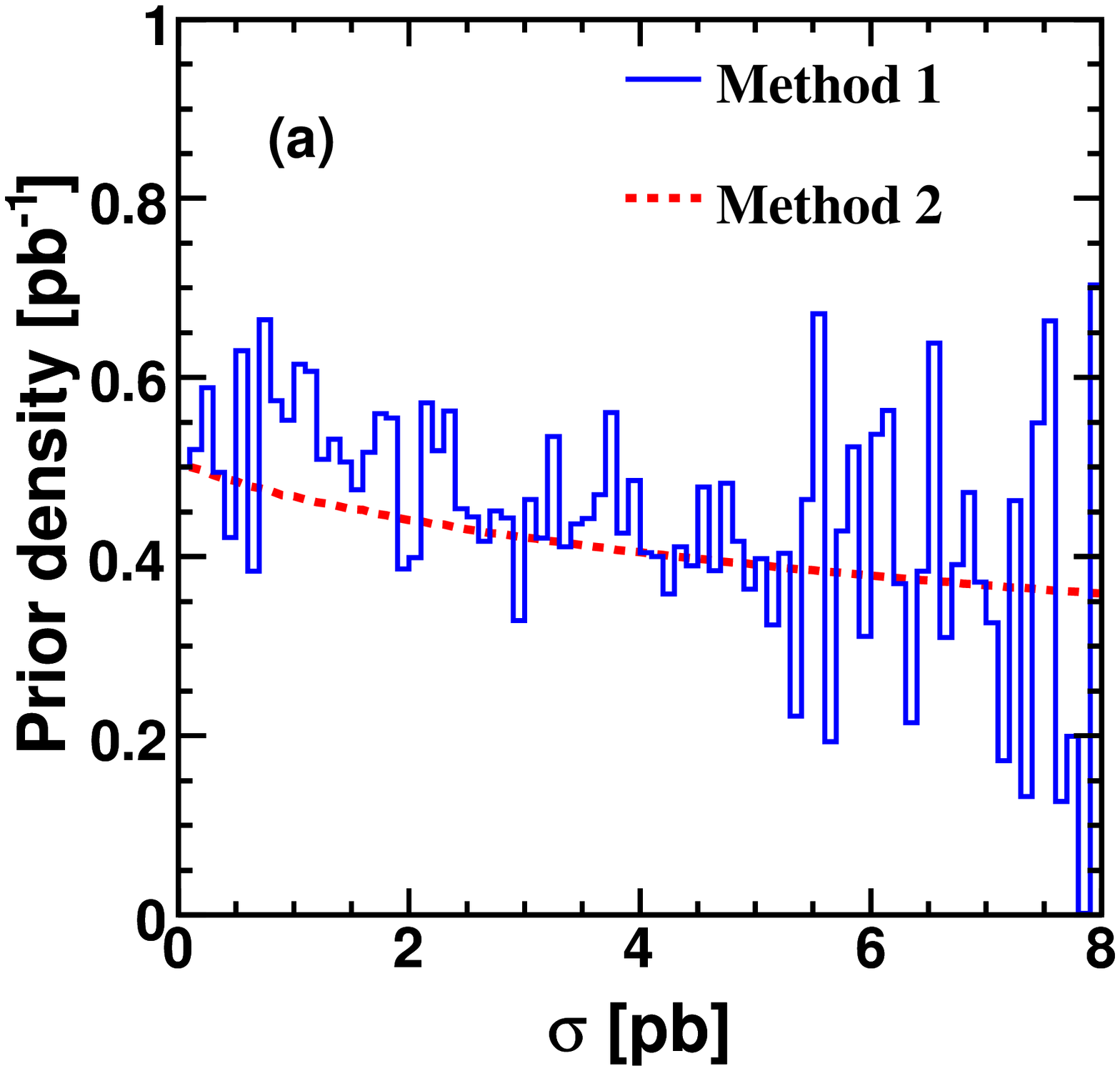}
\hspace*{1cm}
\includegraphics[width=7cm]{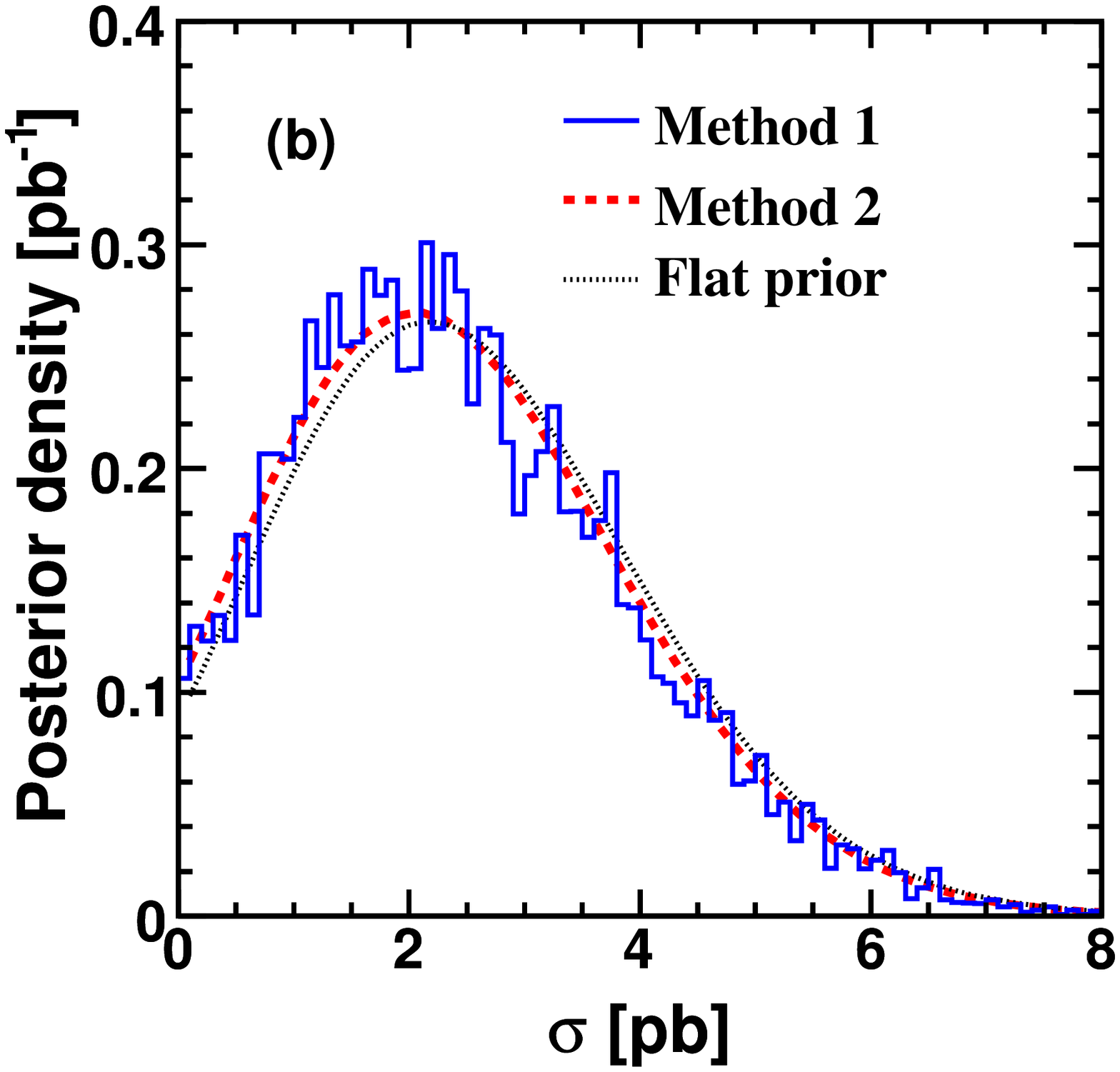}
\caption{(a) Prior densities computed using Method~1 (histogram) and Method~2 
(dashed curve), and (b) corresponding posterior densities, for one of the 
channels in the D0 single top measurement and using 2.3 fb$^{-1}$ of data. For 
comparison we show the posterior density computed using a flat prior (dotted 
curve).
\label{fig:d0results}}
\end{figure}

We model information about the effective integrated luminosity $\epsilon$ and 
the background $\mu$ for each bin with the help of the gamma priors of 
Eq.~\eqref{eq:emuprior}.  These evidence-based priors describe the uncertainty 
due to the finite statistics of the Monte Carlo simulations.  We do not include
systematic uncertainties in this example.  Figure~\ref{fig:d0results}(a) shows 
a comparison of the reference prior for the cross section using Methods 1 (the 
histogram) and 2 (the dashed curve).  The jaggedness of the Method 1 prior 
reflects the fluctuations due to the Markov chain Monte Carlo~\cite{BAT} 
sampling of the parameters.  The increased jaggedness at large $\sigma$ is due
to the fact that the numerical algorithm samples from the flat-prior posterior, 
whose density rapidly decreases in this region.  It is noteworthy that the 
priors computed using the two methods are very similar for this particular 
example.  This is also reflected in the similarity of the posterior densities, 
shown in Fig.~\ref{fig:d0results}(b).  In principle fluctuations in the 
calculated posterior can be made arbitrarily small by increasing the size of 
the Monte Carlo sample.  For reference, Fig.~\ref{fig:d0results}(b) also shows 
the posterior density using a flat prior for the cross section.  An obvious 
conclusion can be drawn: when the dataset is large, here of order 500 events, 
the precise form of the reference prior is not important.  However, for small 
datasets --- which is typical of searches for new or rare phenomena, one should 
expect the form of the prior to matter.  It is then important to use a prior 
with provably useful properties, such as the ones enumerated in 
Sec.~\ref{sec:RefPriors}. 

\section{Discussion}
\label{FinalComments}
Our main purpose in this paper was to propose a set of formal priors with 
properties that make them attractive for use in the analysis of high energy
physics data.  Aside from the theoretical properties of invariance, coherence
and sampling consistency, the reference prior method has three important 
practical advantages: (1) priors can be defined for almost any problem, regardless 
of the complexity of the likelihood function and the number of nuisance and 
interest parameters, (2) in contrast with flat priors, reference priors have so
far always yielded proper posteriors, and (3) reference priors are 
computationally tractable, as shown by the single-top example.

Here we have limited our numerical investigations to the class of likelihood 
functions that are derived from Poisson probability mass functions.  For this 
class the Method-1 reference prior agrees with Jeffreys' rule.  For other 
classes the compact subset normalization argument may introduce a difference.
A possible generalization of our treatment is to unbinned likelihoods.  Since 
our Method-1 and 2 numerical algorithms make no assumptions about the likelihood 
function, they can be generalized to the unbinned case.  However, the Method-1 
algorithm does not implement the compact subset normalization and is therefore 
only applicable to cases where this procedure makes no difference.  Method~2 
requires no compact subset normalization but makes an extra assumption about the 
conditional nuisance prior.

For problems that involve a single continuous parameter or that can be reduced
to this case by a Method-2-type integration, Ref.~\cite{Bernardo2005} 
proposes a numerical algorithm that is based directly on 
Eq.~\eqref{eq:refformula} and is therefore very general.  However we found 
that this algorithm presents some difficulties for the complicated likelihood 
functions used in high energy physics.  One difficulty is the round-off error 
in the product of large numbers of probability densities.  Another difficulty 
is the assessment of the convergence of the integrals in the formula, and of 
the convergence of the finite-sample priors to the reference prior.

Another possible generalization is to problems with more than one parameter of 
interest, as for example in the measurement of the individual single top 
production cross sections in the $s$ and $t$ channels.  For this situation the 
reference prior algorithm requires one to sort the parameters of interest by 
order of importance~\cite{Bernardo2005}, and the results may depend on this 
ordering.  A possible interpretation of this dependence is that it is a measure 
of the robustness of the result to the choice of prior.  This is an area that 
requires more study.

We have developed a general-use software package that implements the methods 
described in this paper and have released it to the Physics Statistics Code 
Repository (phystat.org). 

Finally, we note that the main ideas underlying the construction of reference
priors, namely generality, reparametrization invariance, coherence, and
sampling consistency, have motivated the development of methods for 
summarizing reference posteriors via point estimates, intervals, and hypothesis
tests.  This subfield of objective Bayesianism is known as reference 
analysis~\cite{Bernardo2005,Demortier2005}.

\begin{acknowledgments}
We thank the D0 Collaboration for its support of this work and for granting us 
permission to use a channel of its single top data.  We also thank the members 
of the CMS Statistics Committee for many useful discussions. This work was 
supported in part by the U.S. Department of Energy under grant 
Nos. DE-FG02-91ER40651, DE-FG02-04ER41305, and DE-FG02-95ER40896.
\end{acknowledgments}




\begin{thebibliography}{99}
\bibitem{FormalPriors}
   There is no agreement on what to call such priors.  In the literature one 
   finds many names: objective, non-subjective, non-informative, 
   minimally-informative, default, neutral, conventional, reference.  However, 
   we prefer the term formal prior because it reflects the manner in which they 
   are constructed.
\bibitem{2derivative}
   In our calculations we have used the approximation 
   $d^2f/d\sigma^2 \approx [4 D(f,h)/3 - D(f,2h)/12]/h^2 + \bigcirc(h^4)$, where
   $D(f,h) = f(\sigma+h) - 2f(\sigma) + f(\sigma-h)$ and $h$ is the step size.
\bibitem{MultipleBinFlatPrior}
   The improper posterior pathology mentioned in the introduction disappears
   when data are collected in more than one channel or bin.  The CDF and D0 
   single-top analyses are therefore unaffected by this problem.
\end{thebibliography}
\end{document}